\documentclass[sn-mathphys,Numbered]{sn-jnl}% Math and Physical Sciences Reference Style
\usepackage{graphicx}%
\usepackage{multirow}%
\usepackage{amsmath,amssymb,amsfonts}%
\usepackage{amsthm}%
\usepackage{mathrsfs}%
\usepackage[title]{appendix}%
\usepackage{xcolor}%
\usepackage{textcomp}%
\usepackage{manyfoot}%
\usepackage{booktabs}%
\usepackage{algorithm}%
\usepackage{algorithmicx}%
\usepackage{algpseudocode}%
\usepackage{listings}%
\usepackage{bm}
\usepackage[shortcuts]{extdash}
\usepackage{ragged2e}
\usepackage{textcomp}
\usepackage{lmodern}

\newcommand{\bettershortstack}[2][c]{%
  \begin{tabular}[b]{@{}#1@{}}#2\end{tabular}%
}
\newcommand{\erf}{\ensuremath{\operatorname{erf}}}

\begin{document}

\title[Article Title]{\centering Spectroscopy of VUV luminescence \\ in dual-phase xenon detectors}


\author*[]{\fnm{K.C.} \sur{Oliver-Mallory}}\email{k.oliver-mallory@imperial.ac.uk}

\author[]{\fnm{A.M.} \sur{Baker}\footnote[2]{Now at King's College London.}}
%\email{albert.m.baker@kcl.ac.uk}

\author[]{\fnm{E.} \sur{Jacquet}}
%\email{elisa.jacquet19@imperial.ac.uk}

\author[]{\fnm{T.J.} \sur{Sumner}}
%\email{t.sumner@imperial.ac.uk}

\author[ ]{\fnm{H.M.} \sur{Ara{\'u}jo}}
%\email{h.araujo@imperial.ac.uk}

\affil[ ]{\orgdiv{Department of Physics}, \orgname{Imperial College London}, \orgaddress{\street{Blackett Laboratory, Prince Consort Road}, \city{London}, \postcode{SW7 2AZ}, \country{UK}}}

%\affil[2]{\orgdiv{Department}, \orgname{Organization}, \orgaddress{\street{Street}, \city{City}, \postcode{10587}, \state{State}, \country{Country}}}

%\affil[3]{\orgdiv{Department}, \orgname{Organization}, \orgaddress{\street{Street}, \city{City}, \postcode{610101}, \state{State}, \country{Country}}}


\abstract{We present spectroscopic measurements of xenon luminescence in a time projection chamber operated in a dual-phase (liquid-gas) configuration. Thorium-228 $\alpha$ decays excited the liquid, resulting in the formation of singlet and triplet excimers that emit vacuum ultraviolet (VUV) scintillation. Ionisation electrons were drifted to the liquid surface and extracted into the vapour, where they produced VUV electroluminescence. A time-resolved photon-counting technique was used to obtain the scintillation spectrum in the liquid, which exhibited a peak wavelength of $177.1\pm0.1_\mathrm{stat} \pm0.1_\mathrm{sys}$\,nm and a full-width at half maximum (FWHM) of $11.3\pm0.2_\mathrm{stat} \pm0.0_\mathrm{sys}$\,nm. The data were also used to obtain distinct singlet and triplet emission models, with the singlet emission peaking $1.8\pm0.3_{\mathrm{stat}}\pm 0.3_{\mathrm{sys}}$\,nm shorter than the triplet. The gas electroluminescence spectrum was obtained simultaneously, while under conditions of thermal equilibrium. It remained consistent across vapour pressures of 1.3--2.2\,bar, with a peak of $173.28\pm0.02_\mathrm{stat} {_{-0.1}^{+0.2}}{}_\mathrm{sys}$\,nm, a FWHM of $10.59\pm0.03_\mathrm{stat} {_{-0.2}^{+0.0}}{}_\mathrm{sys}$\,nm, and a small short-wavelength tail that constitutes $(0.6\pm0.1)$\% of the total spectrum. These are the only spectroscopic measurements of liquid scintillation and gas electroluminescence acquired simultaneously to date, and the first such measurements of singlet and triplet emission in the liquid phase. They are important for precisely characterising dual-phase xenon detectors used to search for dark matter particle interactions and other rare events.}

\keywords{VUV emission spectrum, liquid xenon, liquid noble gases, scintillation, electroluminescence, photon counting, dark matter searches, neutrino detectors, time projection chambers}



\maketitle

%%%%%%%%%%%%%%%%%%%%%%%%%%%%%%%%%%%%%%%%%%%%%%%%%%%%%%%%%%%%%%%%%%%%%%%%%%%%%%%%%%%%
\section{Introduction}\label{sec1}

Xenon is widely used for particle detection, in part because it has excellent scintillation and electron transport properties~\cite{chepel13}, allowing simultaneous measurements of the scintillation and ionization response following particle interactions.

Experiments using several tonnes of liquid xenon (LXe) have set world-leading constraints on the scattering of dark matter (DM) particles~\cite{LZFirst,XeNTFirst,PX4TFirst}, and have found the first evidence of coherent nuclear scattering (CNS) of $^8$B solar neutrinos~\cite{LZB8,XENONnTB8,PandaX4TB8}; a previous detector was used to make the first measurement of $^{124}$Xe two-neutrino double-electron capture~\cite{Xe1TXe124}; a smaller experiment with a $^{136}$Xe-enriched target conducted one of the most sensitive searches for neutrinoless double beta ($0\nu\beta\beta$) decay~\cite{EXO-200}; and a series of experiments aim to observe the hypothetical $\mu\rightarrow e\gamma$ phenomenon~\cite{MEGII,MEG}. There are proposals to build very large detectors, with LXe masses approaching $100$\,tonnes, to search for DM and $0\nu\beta\beta$ decay~\cite{G3XeDM,nEXO,DARWINDM,DARWIN0vBB}, as well as CNS of astrophysical and reactor neutrinos~\cite{G3XeDM,CEvNS_reactor,CEvNS_reactor2}. Additionally, high-pressure gaseous xenon (GXe) technology is being developed to search for the distinct two-electron track signature of $0\nu\beta\beta$~decay that distinguishes it from radioactivity backgrounds~\cite{HPGXeTPC,NEXT}. These experiments employ time projection chambers (TPCs) to detect particle interactions in large volumes of ultrapure cryogenic liquid ($165$--$178$\,K) or room-temperature gas --- the particles can be $\gamma$~rays, electrons, protons, $\alpha$~particles, neutrons, neutrinos, or other (e.g.~dark matter).
%Experiments using several tonnes of liquid xenon (LXe) have set world-leading constraints on the scattering of dark matter (DM) particles~\cite{LZFirst,XeNTFirst,PX4TFirst}; a previous detector was used to make the first measurement of $^{124}$Xe two-neutrino double-electron capture~\cite{Xe1TXe124}; and a smaller experiment with a $^{136}$Xe-enriched target conducted one of the most sensitive searches for neutrinoless double beta ($0\nu\beta\beta$) decay~\cite{EXO-200}. 
%probe neutrino properties beyond the standard model of physics.

A dual-phase TPC configuration is commonly employed to achieve the very low energy thresholds ($\sim\!$keV) required to search for dark matter scattering. In such instruments, a massive liquid phase is topped by a thin layer of vapour in thermal equilibrium, which is used to detect the ionisation component of energy depositions in the liquid. Interactions in the liquid target generate localised atomic excitations that rapidly lead to primary scintillation (commonly termed the S1 signal). They also liberate ionisation electrons that are drifted to the liquid surface and are extracted into the vapour ($1$--$2$\,bar) by an applied electric field. Once in the gas, the electrons are accelerated to energies at which they can excite xenon atoms in their path, resulting in electroluminescence (traditionally called secondary scintillation, hereafter referred to as the S2 signal). This emission has the same basic production mechanism as the S1 signal, except that it is always stimulated by electron impact, and the xenon density will be vastly different in the gas phase compared to the liquid. The electric field in the vapour is typically $5$--$10$~kV/cm, which enables efficient electron extraction from the liquid surface, but is not so high as to lead to significant charge multiplication. In a dual-phase configuration, both signals are optical in nature and detectors are commonly instrumented with photomultipliers that are sensitive to the vacuum ultraviolet component (VUV) of the S1 and S2 responses.

\begin{figure}[ht]
  \centering
  \includegraphics[width=0.9\textwidth]{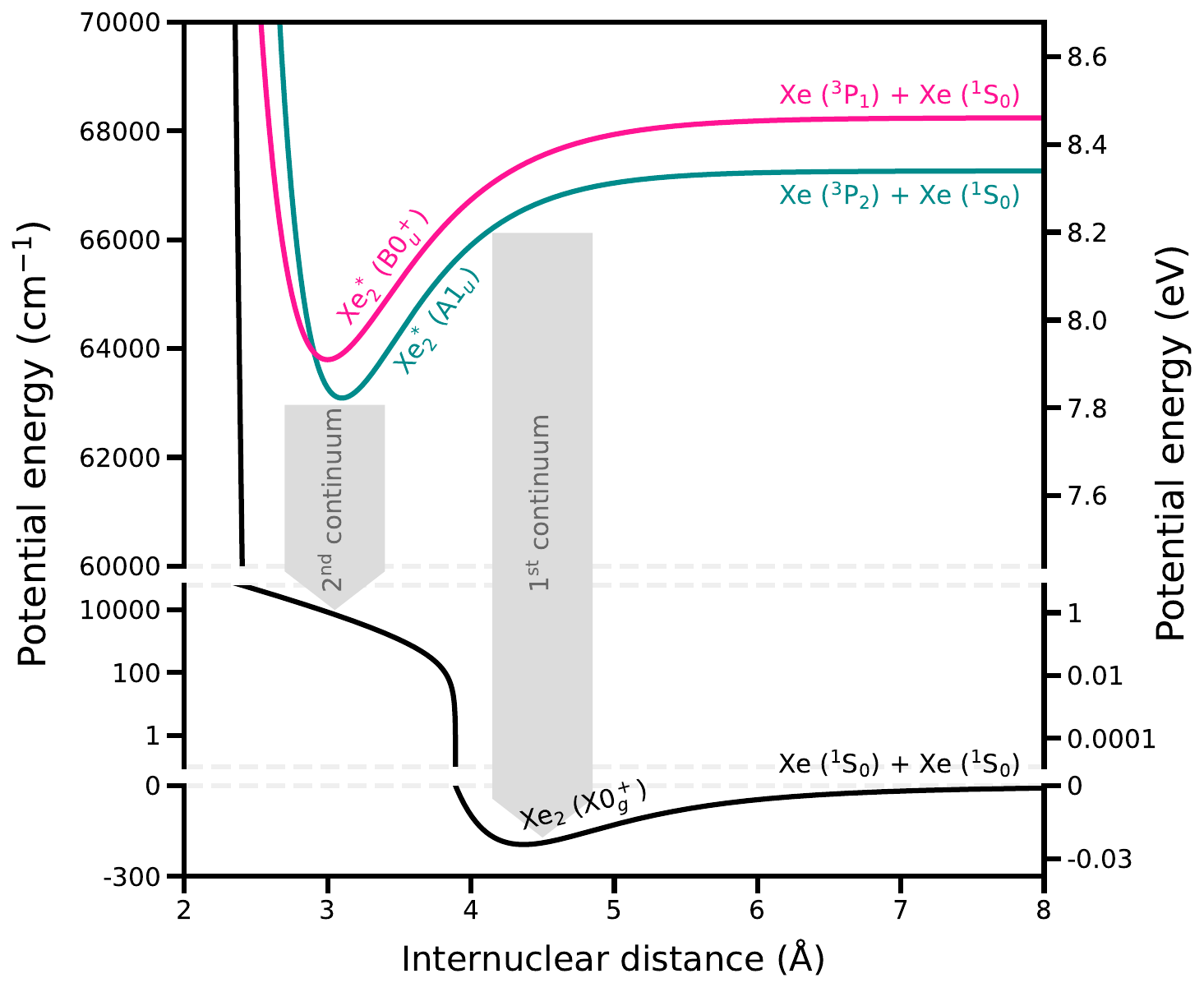}
  \caption{Potential energy curves for the formation of the ground state xenon dimer [Xe$_2$(X$0_g^+$)] and the first two excited state dimers [Xe$_2^*$(B$0_u^+$),~Xe$_2^*$(A$1_u$)]. These are calculations of the potential energy between two xenon atoms in vacuum~\cite{xiaowei2020,nee2000}; the influence of other nearby atoms is neglected, therefore these curves should be considered as an approximation in the liquid and gas phases. The energy regimes of fluorescence from the first and second continua are illustrated by arrows. Note that the low-energy region is plotted in logarithmic scale to reveal the shape of the ground state.}
  \label{fig:xelevels}
\end{figure}

The mechanism for emission of VUV radiation in xenon can be understood in terms of the potential energy diagram for diatomic molecules shown in Fig.~\ref{fig:xelevels}. The limit of large internuclear separation corresponds to isolated xenon atoms, at an infinite distance from each other, in their own energy eigenstates. As one atom approaches another, they form a Van~der~Waals molecular dimer with electronic states described by potential energy curves such as those drawn in the diagram. There is a ground-state dimer [Xe$_2$(X$0_g^+$)] with a shallow potential well that has an equilibrium internuclear distance of $\sim\!4.36$\,\AA{} and depth of only $\sim\!24$~meV, which contains some $\sim\!25$ vibrational levels~\cite{XeGroundState}. There are also singlet and triplet excited state dimers [Xe$_2^*$(B$0_u^+$),~Xe$_2^*$(A$1_u$)] associated with the lowest energy atomic excited states [Xe($^3$P$_1$),~Xe($^3$P$_2$)], which have deeper potential wells ($551$,~$517$\,meV) at shorter equilibrium internuclear distances ($3.0$,~$3.1$\,\AA{}), for which there are $\sim\!45$ vibrational levels~\cite{Xe2StarMorsePotential}. Not shown in the diagram are several additional excited states associated with the lowest-energy atomic states which are purely repulsive or have very shallow potential wells, and many other potential energy curves for higher energy excited states and a conduction band~\cite{Formalism1,Formalism2}.
%%%%%%%%%%
%The calculation in Ref.~\cite{XeGroundState} can be used to estimate the percentage of xenon atoms bound in these vibrational states at room temperature (\textcolor{red}{$\bf{X}$}) and at cryogenic LXe temperatures (\textcolor{red}{$\bf{X}$}).
%%%%%%%%%%

Particles interacting in xenon can excite the molecular and atomic ground states to any of the excited states or to a conduction band (i.e. ionisation). A fraction of the ions will recombine with free electrons to form additional excitation. The higher energy excited states relax down to the lowest energy excited states through a variety of non-radiative mechanisms~\cite{FateOfExcitation} and the emission of infrared light~\cite{piotter23,Hammann_2024}. The atomic state Xe($^3$P$_1$) can decay by emission of a $147$~nm photon that is highly absorbed and thus only observed at very low pressures (below a few mbar), while the radiative transition of the metastable state Xe($^3$P$_2$) to the ground state is forbidden by selection rules. As a consequence, relaxation of the medium follows an alternative path, whereby three-body collisions of the excited atoms with neighboring atoms result in formation of the triplet and singlet excimers in high vibrational/rotational states. These can undergo radiative dissociation emitting photons in a ``first continuum'' of wavelengths, around {${\sim\!150}$\,nm~\cite{DecayRates2}}. However, in liquid or high-pressure gas, excimers experience rapid collisional deexcitation toward lower vibrational/rotational states, reaching thermal equilibrium with the surrounding medium. The subsequent radiative dissociation produces emission in a ``second continuum'', centred around {${\sim\!175}$\,nm~\cite{DecayRates2}}. This is the VUV emission detected by the xenon TPCs mentioned above. Other emissions are also present, such as the infrared light from transitions between higher energy electronic~/~vibrational states~\cite{piotter23,Hammann_2024} and a weak visible-region continuum from neutral bremsstrahlung~\cite{henriques22}, but VUV emission is the main mechanism exploited by xenon detectors.
%In the case of thermal equilibrium between the excimers and the surrounding medium, the occupancy of the excimer vibrational states will depend on temperature through Boltzmann statistics.

High-precision measurements of the S1 and S2 emission spectra are important for xenon experiments: they inform optical models used to derive the absolute response of a detector to primary scintillation and ionisation quanta, and they help understand unusual signal topologies that constitute backgrounds in rare event searches. They are important because small differences in wavelength within the VUV regime can correspond to noticeable changes in optical parameters. For example, the Rayleigh scattering length in LXe varies from $20$--$40$\,cm over a range of $170$—$180$\,nm, calculated from the model and refractive index in Ref.~\cite{seidel02,refractive_index_liq2}; the quantum efficiency (QE) of Hamamatsu R11410 photomultipliers --- used in several such experiments --- rises from zero to $\sim\!40$\% over the short-wavelength side of the second continuum~\cite{R11410QE}; there are variations in the reflectivity of detector surfaces~\cite{PTFERef,PTFEref2}, including at the liquid-phase boundary from changes in the refractive index of the liquid~\cite{refractive_index_gas,refractive_index_liq1,refractive_index_liq2}; and different photoelectron yields can arise from VUV-induced photoionisation of impurities in the liquid or from photoelectric emission from detector surfaces. Rigorous optical models are needed for the precise understanding of data from the current generation of experiments, as well as for the design of future ones.
%It it important to know the S1 and S2 emission spectra with high precision: to inform the optical models used to derive the underlying response of xenon detectors in terms of primary scintillation and ionisation quanta, and to help understand unusual signal topologies which can constitute backgrounds in rare event searches.
%in the VUV regime

%Optical models are used as inputs in calibrations of xenon detectors and in design work for future experiments. To reach percent-level precision, they need to sample from photon spectra with equal or greater precision.
%Optical models of xenon detectors with percent-level precision require the same
%design detectors for
%- high light collection efficiency
%- high energy resolution
%- perform LCE corrections

\vspace{5mm}

Studies of xenon scintillation in the liquid phase have produced several measurements of the second continuum that are not entirely consistent with each other. A high-resolution spectral measurement made under $\gamma$-ray irradiation of cryogenic liquid found a peak wavelength of $174.8\pm0.1_{stat}\pm0.1_{sys}$\,nm and a FWHM of $10.2\pm0.2_{stat}\pm0.2_{sys}$\,nm~\cite{Fujii}. An earlier investigation observed a markedly longer peak wavelength of $178.0\pm 0.6$\,nm, and a wider FWHM of $14.2\pm 1.6$\,nm~\cite{Jortner}, using $^{210}$Po $\alpha$ decays to excite the liquid medium. A study with Xe($^1$S$_0$)$\rightarrow$Xe($^3$P$_1$) selective excitation near the critical point ($290$\,K,~$58.4$\,bar) reported a peak that agrees with Ref.~\cite{Fujii}, but observed a broader spectrum with a FWHM of $13.3\pm0.2$\,nm~\cite{Wahl}.
%$5p^6\rightarrow 5p^56s$

Several factors should be considered when analysing these discrepancies. First, some of the liquid studies may have been affected by trace contamination as they employed less effective purification techniques~\cite{Fujii} or did not purify the xenon beyond supplier specifications~\cite{Jortner,Wahl}. An impurity, such as O$_2$, that absorbs across the short-wavelength side of the spectrum would shift the measurement towards longer wavelengths~\cite{Lu}, which could account for the observed difference in the peaks of the two cryogenic liquid measurements~\cite{Fujii,Jortner}. We also note that the higher temperatures of the study near the critical point~\cite{Wahl} would result in a greater proportion of excimers in higher-energy vibrational~/~rotational states, which should broaden the spectrum and shift it toward shorter wavelengths~\cite{FranckCondon} compared to the cryogenic measurements. A particularly important consideration is that each study used a different method to excite the xenon, which would yield different ratios of singlet to triplet excimers. Significantly, $\alpha$ decays produce a much greater ratio of singlet-to-triplet emission ($\sim\!0.45$ in LXe~\cite{T1T3A}) than electrons and $\gamma$ rays ($\sim\!0.05$~\cite{LUXPSD,T1T3C,T1T3E}), and there are indications that selective excitation to either Xe($^3$P$_1$) or Xe($^3$P$_2$) would result primarily in triplet emission because of an efficient transfer from the high vibrational levels of the singlet to the triplet state~\cite{Marchal}. An investigation of the separate emissions in room-temperature gas found a $\sim\!1$\,nm longer peak wavelength for the triplet, along with a $\sim\!1$\,nm narrower FWHM~\cite{SingletTriplet}. However, this does not appear to explain the discrepancy observed between the two cryogenic liquid studies, as the measurement with a higher triplet fraction showed a shorter peak wavelength~\cite{Fujii,Jortner}. It is clearly beneficial to conduct similar measurements in high-purity cryogenic liquid.

Regarding gas-phase luminescence, there is only one spectrocopic measurement in the pressure and temperature regime of dual-phase TPCs, and this examined prompt scintillation caused by $\gamma$-ray irradiation rather than a genuine electroluminescence signal. As reported in Ref.~\cite{Murayama}, the study was carried out with xenon gas at $1.4$\,bar and $174$\,K, finding emission in the second continuum with a peak wavelength of $173.4\pm3$\,nm and FWHM of $14.4$\,nm. There exist also $\alpha$-decay, selective-excitation, and electron-impact measurements in room-temperature gas of $0.5$--$34$\,bar resulting in peak wavelengths and FWHMs ranging from $169$--$173$\,nm and $11$--$15$\,nm~\cite{Jortner,Wahl,Koehler,Suzuki}. However, the electron-impact studies were performed in the energy regime where electrons can ionise xenon atoms, which differs from that of typical electroluminescence where electrons can excite, but not ionise, the medium.
%, and no obvious contribution from the first continuum.
%Other measurements in room temperature gas were made using $\alpha$ decays ($1$~bar) or selective excitation ($40$~bar), finding peak wavelengths and FWHMs of $175$/$172$\,nm and $15$/$13$\,nm.

There is a consistent overall trend in the liquid and gas measurements, with all gas-phase results exhibiting a shift toward shorter wavelengths, regardless of the temperature, pressure, or method used to excite the xenon. While there may be unaccounted for systematic effects that coincidentally contribute to this trend, such an explanation would not account for a similar observation made with a single experimental setup~\cite{Wahl}, where an abrupt $\sim\!3$\,nm increase in wavelength was recorded when cooling xenon from the gas to the liquid phase near the critical point. This shift between phases could result from changes to the excimer potential energy surfaces caused by interactions with the surrounding medium. Multiparticle interactions are neglected in the binary potential energy curves presented in Fig.~\ref{fig:xelevels}, which may provide a reasonable approximation for the gas phase, depending on the pressure, but are likely insufficient to fully describe the conditions in the liquid. Changes in atomic number density within a single phase also influence the emission spectrum, though to a lesser extent than the shift observed between phases~\cite{Wahl,Koehler}. Additionally, it is possible that gas-phase excimers decay prior to achieving complete thermal equilibrium with their environment, even at higher pressures. In $\sim\!1$\,bar xenon gas, the interval between atomic collisions (several-hundred-picoseconds) is not much shorter than the decay times of the singlet and triplet excimers ($2$--$5$ and $21$--$32$\,ns~\cite{LUXPSD,T1T3A,T1T3B,T1T3C,T1T3D,T1T3E,T1T3F}). In contrast, the collision interval in the liquid phase is significantly shorter, on the order of picoseconds. Clearly, it is valuable to perform simultaneous measurements of gas and liquid emission spectra under the actual operating conditions of dual-phase TPCs, as this allows for more effective control of systematic effects and enables detection of subtle variations in the gas emission spectrum with pressure.

In this work, a LXe-TPC has been constructed and used to obtain high-resolution emission spectra of the S1 and S2 signals simultaneously, employing a time-resolved photon counting technique with a VUV monochromator. The measurements were performed using high-purity xenon at equilibrium vapour pressures that span the operating range of current dark matter search experiments that use this technology ($1.3$, $1.7$, and $2.2$\,bar). For the first time, we are able to compare the two emission spectra under common and controlled conditions, limiting possible systematic effects that could have led to inconsistencies between past measurements. We also perform a first measurement of the separate singlet and triplet emission spectra in the cryogenic liquid.
%, purified with a hot zirconium getter,

%%%%%%%%%%%%%%%%%%%%%%%%%%%%%%%%%%%%%%%%%%%%%%%%%%%%%%%%%%%%%%%%%%%%%%%%%%%%%%%%%%%%%%%%%
\section{Experimental setup}\label{sec2}
\begin{figure}[ht]
  \centering
  \includegraphics[width=\textwidth]{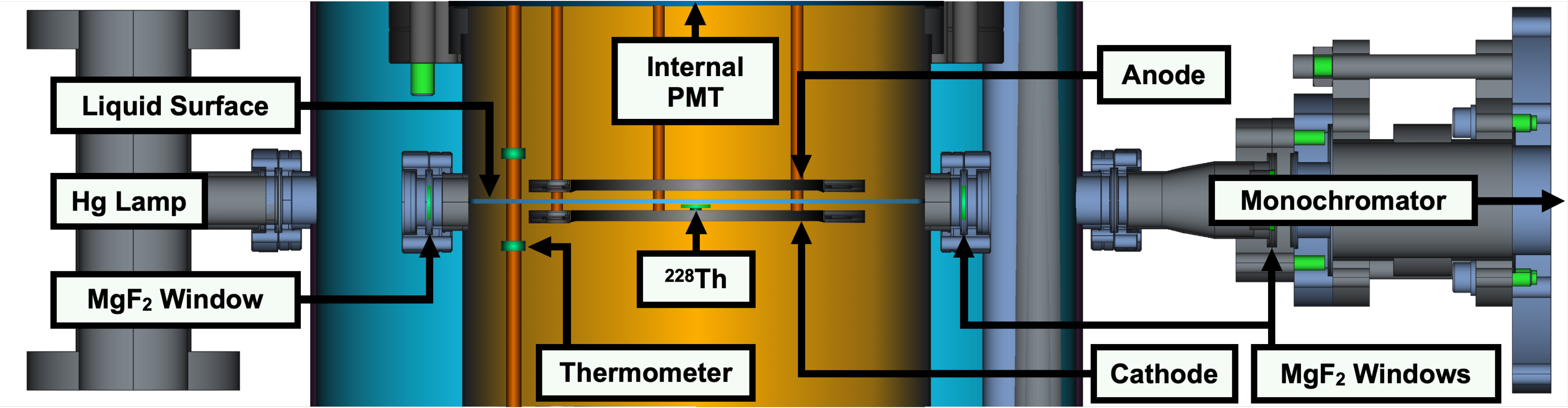}
  \caption{An illustration of the primary components of the optical system and simplified TPC. Xenon luminescence was detected either directly by the internal PMT, or transmitted through two MgF$_2$ viewports and a monochromator to a secondary spectroscopy PMT (not depicted). For calibration purposes, a low-pressure Hg arc lamp was strategically positioned directly opposite the monochromator, ensuring that its emitted light followed the same optical path as the xenon luminescence.}
  \label{fig:CAD}
\end{figure}

The optical measurement was conducted with a simplified LXe TPC contained in a vacuum-insulated cryostat, as illustrated in Fig.~\ref{fig:CAD}. Some $6$\,kg of liquid xenon were condensed into a temperature-controlled inner vessel (IV) so that the surface lay approximately midway between a set of cathode and anode parallel-wire electrodes, spaced $13$\,mm apart and biased to a potential difference of $9$~kV. A $\sim\!1$\,kBq $^{228}$Th radioactive source deposited on a platinum disc was attached to the centre of the cathode. This isotope has a decay chain with six $\alpha$ emissions ($5.5$--$9$\,MeV) that excited the liquid just above the disc, resulting in primary scintillation (S1 pulse). The $\alpha$ particles also generated ionisation electrons that were extracted from the liquid by a uniform vertical electric field; once in the vapour, the drifting electrons generated electroluminescence photons (S2 pulse). The liquid below this partial TPC structure played no appreciable role in the measurement. 
%The IV ($15$\,cm diameter $\times$ $19$\,cm length) was filled with $\sim\!6$\,kg of liquid so that the surface lay approximately midway between a set of cathode and anode parallel wire electrodes, spaced $13$\,mm apart.
% pointing from the anode to the cathode

\subsection{Optical system}\label{subsec2.1}
A vacuum monochromator (Acton~VM-502~\cite{Directindustry} with a $1200$\,grooves/mm Al$+$MgF$_2$ grating) was coupled to an evacuated outer vessel (OV), allowing xenon luminescence photons to pass through two MgF$_2$ viewports to reach its input port, aligned with the liquid surface. The setup enabled simultaneous analysis of both S1 and S2 signals produced just below and above the liquid surface, respectively. Both the OV and the monochromator were pumped to high vacuum. 

\begin{figure}[ht]
  \centering
  \includegraphics[width=0.75\textwidth]{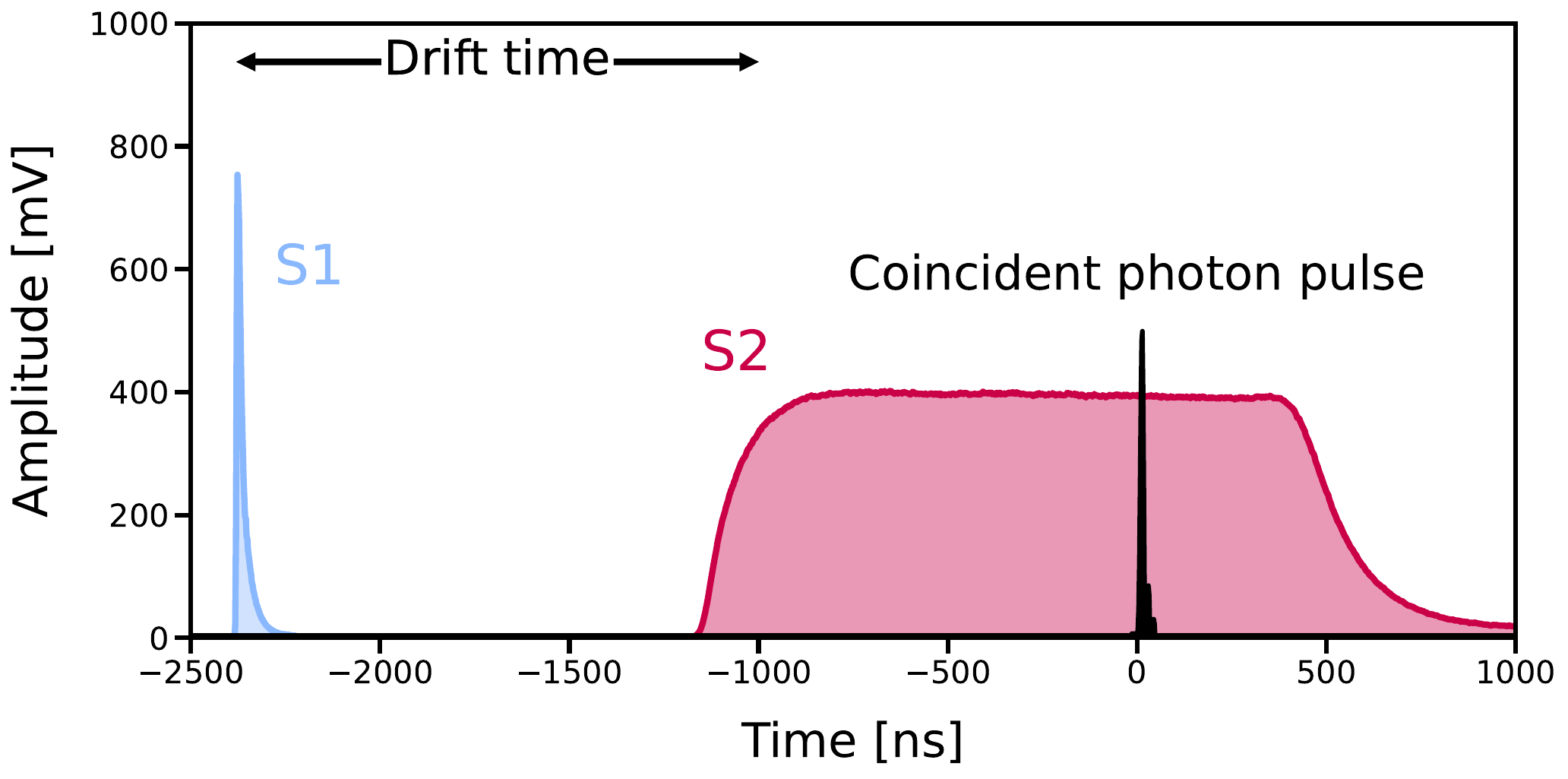}
  \caption{A set of averaged $\alpha$-decay waveforms in the internal PMT (blue and red pulses) together with a single photon pulse (black) from the spectroscopy PMT, detected during the dataset with liquid and vapour in equilibrium at $2.2$\,bar.}
  \label{fig:waveform}
\end{figure}

Measurements were conducted by counting coincident pulses in two photomultiplier tubes (PMTs). An ``internal'' PMT (2-inch quartz-windowed ETL~D730~\cite{PMTs}) detected hundreds-of-thousands of S1 and S2 photons per $\alpha$ decay and was used to map the time profiles of events to identify S1 and S2 pulses, shown in Figure~\ref{fig:waveform}. A fraction of the photons passed through the two MgF$_2$ viewports to the monochromator, which directed a narrow range of wavelengths to a second ``spectroscopy'' PMT (1.13-inch MgF$_2$-windowed ETL~9406B~\cite{PMTs}). These individual photons were detected in coincidence with the S1s or S2s seen by the internal PMT.
%These individual photons were detected in coincidence with the S1s or S2s of the internal PMT. Figure~\ref{fig:waveform} shows a set of averaged waveforms in the internal PMT, and an example detection in the spectroscopy PMT occurring in coincidence with the S2 pulse.
%, one inside the IV and the second coupled to the setup through a monochromator.

The spectroscopy PMT was biased to high gain to produce prominent single photoelectron (SPE) pulses ($\sim\!60$\,mV peak amplitude before amplification) and hence trigger the data acquisition system with high efficiency. It had been our intention to trigger on the internal PMT instead, but the low optical coupling efficiency for the spectroscopy measurement meant that this produced too large a data rate (on average, one photon is detected per $\sim\!10$ $\alpha$-decay events); we confirmed that this alternative trigger did not bias our measurement.
%The voltage applied to the spectroscopy PMT ($-1450$\,V) was chosen to produce prominent single photoelectron (SPE) pulses ($\sim\!60$~mV before amplification), while the voltage applied to the internal PMT ($-800$ to $-1200$\,V) was varied with the size of S2s observed under each set of operating conditions.

The TPC construction is deliberately quite minimal in terms of materials, consisting mostly of stainless steel components held together by PEEK/PTFE supports and HV cabling. The orientation of the monochromator was chosen to maximise the number of scintillation photons reaching its input port while ensuring that the plastic components were outside its field of view --- to avoid the possibility of detecting any faint visible-wavelength luminescence when excited by xenon light~\cite{eBKG,PTFEIntegratingSphere,PMMA&PTFELumin,PMMA&TPBFluor}. 

For spectral calibration, a low-pressure Hg arc lamp~\cite{childs62} was mounted on the OV directly opposite the monochromator, such that its light would travel through the IV and then follow the same path as the xenon scintillation photons. The $184.9$\,nm atomic calibration line (detected FWHM of $\sim\!1$~nm) is much narrower than the xenon luminescence and serves as a good reference signal. 
%The orientation of the monochromator was chosen to maximise the number of scintillation photons reaching its inlet while ensuring that the PEEK/PTFE supports of the TPC were outside of it's field of vision --- plastics like these are hypothesized to produce faint visible-wavelength background scintillation when excited by xenon luminescence~\cite{eBKG,PTFEIntegratingSphere,PMMA&PTFELumin,PMMA&TPBFluor}. For calibration, a low-pressure Hg arc lamp~\cite{childs62} was installed directly opposite the monochromator so that its light would travel through the vessel following the same path as the xenon scintillation.

\begin{figure}[ht]
  \centering
  \includegraphics[width=0.7\textwidth]{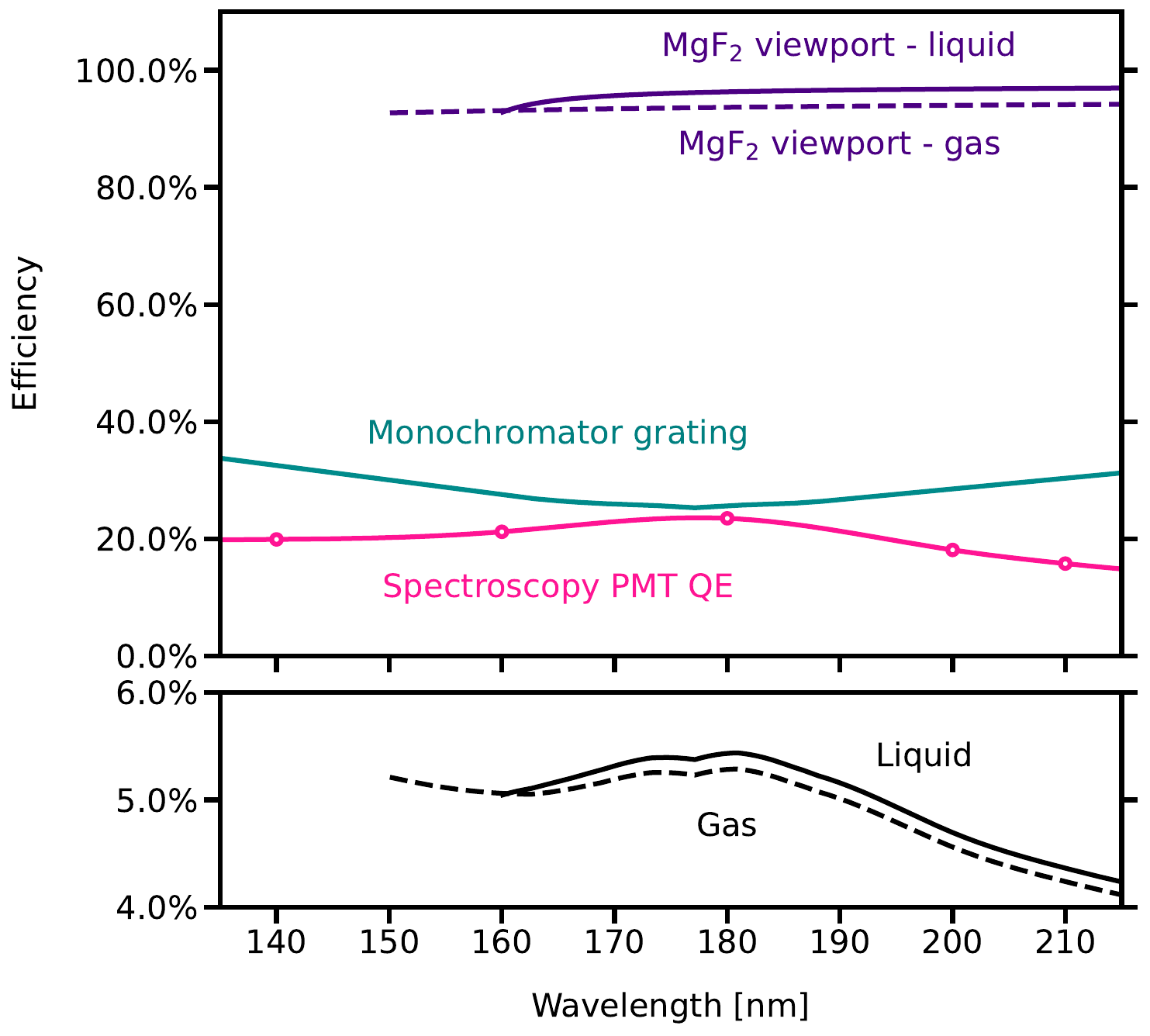}
  \caption{Wavelength-dependent efficiencies of the optical components within the setup. Transmittance calculations for the innermost MgF$_2$ viewport were conducted separately for S1 and S2 photons propagating from liquid and gaseous xenon into MgF$_2$, respectively. These were restricted to wavelengths above $160$ and $150$\,nm, corresponding to the spectral region where significant VUV luminescence was observed. The results were combined with the efficiency of the monochromator grating~\cite{Fujii} and the quantum efficiency of the ETL~9406B PMT~\cite{PMTs}, as shown in the bottom panel. The transmittance of the second MgF$_2$ viewport, which separates the outer vessel from the monochromator, exhibits minimal wavelength dependence and is, therefore, not shown for clarity.}
  \label{fig:efficiency}
\end{figure}

The wavelength-dependent detection efficiency for VUV photons within our setup is depicted in Figure~\ref{fig:efficiency}. This encompasses the transmittance of the MgF$_2$ viewports, the efficiency of the monochromator grating, and the quantum efficiency (QE) of the spectroscopy photomultiplier tube (PMT). The transmittance of the innermost MgF$_2$ viewport was calculated separately for S1 photons originating in the liquid and S2 photons originating in the gas, using the refractive indices from Ref.~\cite{refractive_index_liq2,refractive_index_gas,MgF2_refractive_index}. The monochromator grating efficiency was obtained from Ref.~\cite{Fujii}, which provides data within the wavelength range of of $160$ to $190$ nm. This range encompasses most of the region of interest, and for wavelengths outside this range, the efficiency was estimated through linear extrapolation. The QE data were measured for a different PMT unit of the same model by the manufacturer~\cite{PMTs}, and a spline interpolation was used to estimate the QE at wavelengths between the measured datapoints.
%Figure~\ref{fig:efficiency} presents the wavelength-dependent detection efficiency for VUV photons in our setup, considering the transmittance of the MgF$_2$ viewports, the efficiency of the monochromator grating, and the quantum efficiency (QE) of the spectroscopy PMT. 

The primary sources of uncertainty affecting the overall efficiency stem from the transmittance of the MgF$_2$ viewports and the QE of the PMT. An experimental measurement of the viewport transmittance in vacuum deviates from the theoretical calculation~\cite{Viewports}, presumably due to the light absorption and scattering phenomena. Specifically, the measurement indicates a $\sim\!10$\% reduction in transmittance on the shorter wavelength side of the spectrum compared to the longer wavelength side. Regarding the PMT QE, we lack data on the variability of this parameter across similar samples from the specific manufacturer. However, we have quantified the variability of similar bialkali photocathodes from an alternative supplier~\cite{Hamamatsu} as a $\sim\!14$\% tilt of the QE function within the VUV regime. The impact of these uncertainties on the shape of the S1/S2 spectra is addressed in Sec.~\ref{subsec3.3.1}.
%The primary source of uncertainty on the overall efficiency comes from the QE, and its impact on the shape of the S1/S2 spectra is addressed in Sec.~\ref{subsec3.3.1}.
% The primary source of uncertainty on the overall efficiency comes from the QE, \textcolor{blue}{\textbf{which we have estimated from...}}. The impact of this uncertainty on the shape of the spectra is addressed in Sec.~\ref{subsec3.3.1}.
%which we have estimated from the range of typical values given by the manufacturer
%The QE data points were derived from the cathode radiant sensitivity of a different PMT of the same model, and a spline fit was applied to estimate this parameter at intervening wavelengths. 

%The transmittance of the second MgF$_2$ viewport, which separates the outer vessel from the monochromator, exhibits minimal wavelength dependence and is therefore not shown for clarity.

\subsection{Data acquisition system}\label{subsec2.2}
Signals from the PMTs were fed through $\times10$ broadband amplifiers (Phillips Scientific Model~770) then digitised at $500$\,MS/s with 8-bit resolution by an Acqiris~CC108 crate equipped with two DC270 modules~\cite{ACQIRIS}. The spectroscopy PMT signal was routed through a $70$\,MHz low-pass filter before amplification to suppress electromagnetic interference. The digitisers were triggered by this signal (in Fig.~\ref{fig:waveform} the coincident pulse appears near zero time, corresponding to the trigger position).
%%%%%%%%%%
%The trigger efficiency for single photon pulses was estimated to be \textcolor{red}{\textbf{X}}\%.
%%%%%%%%%%

The raw data were analysed offline, with the PMT waveforms scanned for pulses which were then parametrised and databased for further analysis. An extensive set of pulse-, waveform-, event- and run-level parameters was thus generated, including to detector conditions and monochromator data. 
%MiDAQ, a custom Linux-based software built for the experiment described in Ref.~\cite{MIGDAL}, interfaced with the Acqiris crate to set parameters such as the trigger threshold and sampling rate. For most datasets, we triggered on the spectroscopy PMT using a threshold of \textcolor{red}{$\bf{X}$}\,mV, corresponding to an efficiency for triggering on single photon pulses of \textcolor{red}{$\bf{X}$}\%. The sampling rate was set to $2$\,ns to accurately resolve these pulses ($\text{FWHM}\sim\!10$\,ns) and their start time with respect to S1s and S2s. MiDAQ stored the PMT waveforms in binary format, then a C$++$-framework named MiDAS \cite{MIGDAL} was used offline to reduce the data to pulse-level information such as pulse area, start time, etc.

The narrow range of photon wavelengths at the output of the monochromator were controlled by a stepper motor rotating the diffraction grating inside the instrument. The grating was scanned from a home position corresponding to a wavelength well below the xenon VUV luminescence, and excellent reproducibility was achieved across multiple scans. In each scan the grating was moved by a set number of motor steps ($\Delta \lambda \sim\!1/30$\,nm), dwelling for $5$\,seconds in each position until reaching the end of the scanning range. A single scan took typically several hours, with a run at a fixed set of detector conditions consisting of several such scans in sequence.

\subsection{Xenon handling system}\label{subsec2.3}
The xenon gas used in the experiment was purified prior to liquefaction within the detector. The gas handling system supplying the LXe chamber includes a double-diaphragm pump (KNF~N143.12E-E268~\cite{KNFGroup}) that feeds a heated zirconium getter (SAES~MonoTorr~PS4-MT3-R-2~\cite{SAESGetters}) positioned inline with the IV and a large ($2100$\,L) containment vessel. At the start of operations, the system was evacuated to a residual pressure of $\sim\!10^{-8}$\,mbar before introducing $\sim\!15$\,kg of pure xenon gas at ambient temperature. To purify the gas stock and desorb atmospheric gases from the detector surfaces, the xenon was circulated through the purification loop at the optimal flow rate of the getter for four days, before cryopumping the gas back into the storage cylinders. The chamber construction avoids high-outgassing materials to a large degree, and the LXe remains pure once in the purged chamber --- obviating the need for continuous purification during operation.

Ahead of the data collection run, the chamber was cooled to $\sim\!175$\,K using a helium cryocooler coupled thermally to a copper baseplate at the bottom of the IV; the temperature of this baseplate provides the control point for the system, and hence of the vapour pressure above the liquid. The LXe target was condensed over two days, until the liquid level was visible halfway up the MgF$_2$ viewports.
%The xenon handling system is a loop that includes the IV, a $2100$\,L containment vessel, a double diaphragm pump (KNF N143.12E-E268 \cite{KNFGroup}), and a hot zirconium getter (SAES MonoTorr PS4-MT3-R-2 \cite{SAESGetters}). At the start of operations, the system was evacuated to a residual pressure of $\sim\!10^{-8}$\,mbar before introducing $\sim\!15$\,kg of gaseous xenon from a stainless steel storage bottle. To purify the xenon and flush atmospheric gases from the surfaces of the components, the xenon was circulated through the loop at the optimal flow rate of the getter for four days before cryopumping the gas back into the bottle. At the start of the data collection run, the purified xenon was condensed in the IV using a helium cryocooler to cool the copper baseplate of the vessel to liquid xenon temperatures.

A slow control system continuously monitored key pressures and temperatures, with two PT100 thermometers located $\sim\!1$\,cm above and below the electrodes, and another six positioned at different points along the IV. The residual pressures of the monochromator and the OV were also monitored, and a residual gas analyser tracked the H$_2$O partial pressure in the OV vacuum (and for the presence of any xenon leaks into the OV). Throughout the various runs, the monochromator vacuum remained in the range $10^{-5}$--$10^{-7}$\,mbar, and the partial pressure of H$_2$O in the OV varied between $10^{-5}$--$10^{-9}$\,mbar, exceeding our requirement by several orders of magnitude (at $10^{-3}$\,mbar H$_2$O partial pressure, around $1$\% of the xenon scintillation would be absorbed when traveling through the monochromator and OV vacuum spaces).

Once the IV was filled, the purity of the xenon was confirmed by monitoring the stability of the S2 signal over extended periods of time -- electron transport is highly sensitive to the presence of electronegative impurities. The getter's specified performance  of $<1$\,ppb H$_2$O, O$_2$, and other electronegative species greatly exceeds our requirement of $<1$\,ppm H$_2$O, such that at most $1$\% of xenon scintillation would be absorbed when travelling from the source to the edge of the IV. Direct measurement of the purity via the free electron lifetime in such a compact LXe TPC is challenging; however, we estimated a conservative lower limit of $4\,\mathrm{\mu}$s. This corresponds to an O$_2$ concentration of around $50$\,ppb, based on the electron attachment rate constant from Reference~\cite{AttachmentRateConstant}. While this estimate is greater than the getter's specified performance, it is well below our requirement for the measurements described in this study.

\section{Spectroscopic measurements}\label{sec3}
\subsection{Measurement conditions}\label{subsec3.1}
%The first spectroscopy dataset was started immediately after filling the chamber to a liquid height several millimeters above the midpoint between cathode and anode, measured by photographing the meniscus against a steel ruler installed behind one of two viewports parallel to the liquid surface. In the $12$\,hours following the fill, the average temperature of the chamber cooled down to its equilibrium value and the liquid height decreased to the midpoint. The $\sim10$\,K temperature gradient between the liquid surface and baseplate caused xenon near the bottom of the chamber to freeze, decreasing the liquid height due to the $\sim1.2\times$ greater density of the solid phase. For all datasets, we reached equilibrium with the liquid height at the approximately the midpoint (∼ 6.5 mm above the cathode) so that photons produced both below and above the surface could travel a straight path to the inlet of the monochromator.
 
Data were collected at three vapour pressures that span the operating range of modern LXe dark matter experiments~\cite{LZFirst,XeNTFirst,PX4TFirst}. The first dataset, at $2.2$\,bar, took several days, after which the temperature setpoint was reduced to lower the pressure to $1.7$\,bar for the second dataset --- additional xenon was added to maintain the liquid height. This procedure was repeated for the third dataset at $1.3$\,bar, which was terminated early following an electrical discharge that appeared to compromise the liquid purity, greatly reducing the size of S2s. The xenon was then repurified before taking data with $2.2$\,bar room-temperature gas, first with zero electric field to collect gaseous S1 data ($2$\,days), and then with potential difference of $4$~kV across the electrodes to collect S2 data ($6$\,hr). A final S2 dataset was taken with $0.3$\,bar room-temperature gas ($2.5$\,days, $\Delta V = 3$~kV). Table~\ref{tab:conditions} summarises the measurement conditions for each of the datasets.
%All two-phase data were taken with the liquid surface approximately midway between the cathode and the anode, so that photons produced both below and above the surface had a line-of-sight path to the inlet of the monochromator.
%Additional S1 scintillation data were collected with the liquid surface at two different heights to study possible systematic effects due to reflections on the liquid surface (see Sec.~\ref{subsec3.3.1}).

\begin{table}[h]
\caption{Measurement conditions for all datasets. \textbf{Detector:} The reduced electric field in the gas phase was calculated from the other detector parameters in the table, the separation between the cathode and anode ($13$\,mm), and their potential difference ($9$\,kV for liquid and vapour; $4$/$3$\,kV for $2.2$/$0.3$\,bar room-temperature gas), using a liquid xenon dielectric constant from Ref~\cite{Dielectric}. The liquid height was measured with respect to the cathode from photographs of the liquid surface. \textbf{Spectroscopy:} Parameters not listed below are the slit height ($4$\,mm), interval by which the motor stepped through the scanning range ($1/30$\,nm), and the dwell time at each wavelength within the scanning range ($5$\,sec). The livetime per nanometre is the dwell time divided by the wavelength interval and multiplied by the number of scans performed under stable operating conditions.}
\label{tab:conditions}
\begin{tabular}{@{}lccccccccc@{}}
%\begin{tabular*}{\textwidth}{@{\extracolsep\fill}l|ccccccc}
%\toprule
& \multicolumn{4}{@{}c@{}}{Detector} & \multicolumn{5}{@{}c@{}}{Spectroscopy} \\
\cmidrule(lr){2-5} \cmidrule(lr){6-10}
& \bettershortstack[c]{Press.\\(bar)} & \bettershortstack[c]{Temp.\\(K)} & \bettershortstack[c]{Liquid\\height\\(mm)} & \bettershortstack[c]{Reduced\\E-field\\(Td)} & \bettershortstack[c]{Slit\\width\\(mm)} & \bettershortstack[c]{Scanning\\range\\(nm)} & \bettershortstack[c]{Livetime\\per nm\\ (min)} & \bettershortstack[c]{No.\\S1s\\(\#)} & \bettershortstack[c]{No.\\S2s\\(\#)} \\
%\midrule
\hline
\hline
Liquid & $2.15$ & $179$\footnotemark[1] & $6.4$ & $10$ & $0.25$ & $110$--$210$ & $15$ & $879$ & $18,376$\\
\hspace{0.3cm}\& & $1.73$ & $175$\footnotemark[1] & $5.8$ & $12$ & $0.18$ & $145$--$205$ & $92$ & $2,270$ & $60,013$ \\
Vapour & $1.34$ & $170$\footnotemark[1] & $4.7$ & $14$ & $0.18$ & $110$--$210$ & $22$ & $267$ & $12,234$ \\
\hline
\multirow{2}{*}{\shortstack[l]{Gas}} & $2.21$ & $293$\footnotemark[2] & - & $0$, $5.6$\footnotemark[3] & $0.18$ & $145$--$205$ & $42$, $5$\footnotemark[3] & $1,027$ & $8,935$ \\
 & $0.29$ & $296$\footnotemark[2] & - & $33$ & $0.18$ & $135$--$205$ & $52$ & $0$ & $10,054$\\
\hline
% \botrule
\end{tabular}
\footnotetext[1]{Calculated from the Antoine equation and pressure measurement~\cite{antoine}.}
\footnotetext[2]{Average from two sensors inside the chamber.}
% \footnotetext[3]{Scans performed at different liquid levels to study the effect of reflections are not included.}
\footnotetext[3]{The electrodes were powered off for $2$\,days to collect S1 data, then biased for $6$\,hr to collect S2 data.}
% $756$ & $19,475$
% $1,991$ & $64,090$
% $216$ & $13,677$
% $2,767$ & $20,317$
% $0$ & $20,521$
\end{table}

The monochromator slit widths were adjusted to maximise the intensity of detected light whilst achieving an accurate calibration with sub-nm resolution (details given in Section~\ref{subsec3.2}). During the first dataset the slit width was varied between $0.1$, $0.18$, and $0.25$\,mm while keeping the slit height at the recommended maximum of $4$\,mm. The majority of data at $2.2$\,bar were collected with a slit width of $0.25$\,mm, with a reduction to $0.18$\,mm thereafter.

The monochromator was repeatedly scanned over the xenon excimer emission range to monitor for long-term drift of the measurement conditions. Each scan lasted a few hours and the process was repeated for several days to acquire sufficient statistics; hundreds of S1s and tens-of-thousands of S2s were detected in coincidence with a photon in the spectroscopy PMT, with the benefit that individual scans could be cut from the analysis if they occurred during unstable operating conditions.

We monitored a variety of detector parameters continuously to study possible systematic effects and veto periods of instability, such as those that followed changes in the temperature set-point.  In general good stability was obtained, including:
\begin{itemize}
  \item In the majority of the final data, pressure was stable to within $0.3$\%~{\it RMS}, except for the $1.3$\,bar dataset in which the pressure was still settling (an $8$\% effect) following a temperature set-point change with the electrical discharge occurring before the system had reached equilibrium. At equilibrium, pressure variations were proportional to gas temperature changes which correlated with the environmental temperature.
  \item The liquid height was tracked during two-phase operation via the electron drift time between the radioactive source and the liquid surface, which was stable for the $2.2$ and $1.7$\,bar datasets ($0.5$\%~{\it RMS}) but decreased by $55$\% throughout the $1.3$\,bar dataset.
  \item Xenon purity was stable while liquid and vapour were present in the chamber, as evidenced by the stability in the S2 pulse area during the $2.2$ and $1.7$\,bar datasets ($1$\%~{\it RMS}). A $30$\% increase was observed during the $1.3$\,bar dataset, which can be attributed to the increasing distance between the liquid surface and the anode where electroluminescence occurs, as opposed to improving purity. During room-temperature gas-phase operation the inevitable gradual ingress of contaminates from outgassing caused a disproportionately larger effect due to the much smaller quantity of xenon present in the chamber. In the $2.2$\,bar room-temperature data, the S1 area decreased by $4$\% over the $2$\,days with zero applied field. Over the $6$\,hr in which the electrodes were biased, the S2 area was constant. In the $0.3$\,bar data, the S2 area decreased by $55$\% over $2.5$\,days.
\end{itemize}

\subsection{Calibration}\label{subsec3.2}
\begin{figure}[ht]
  \centering
  \includegraphics[width=\textwidth]{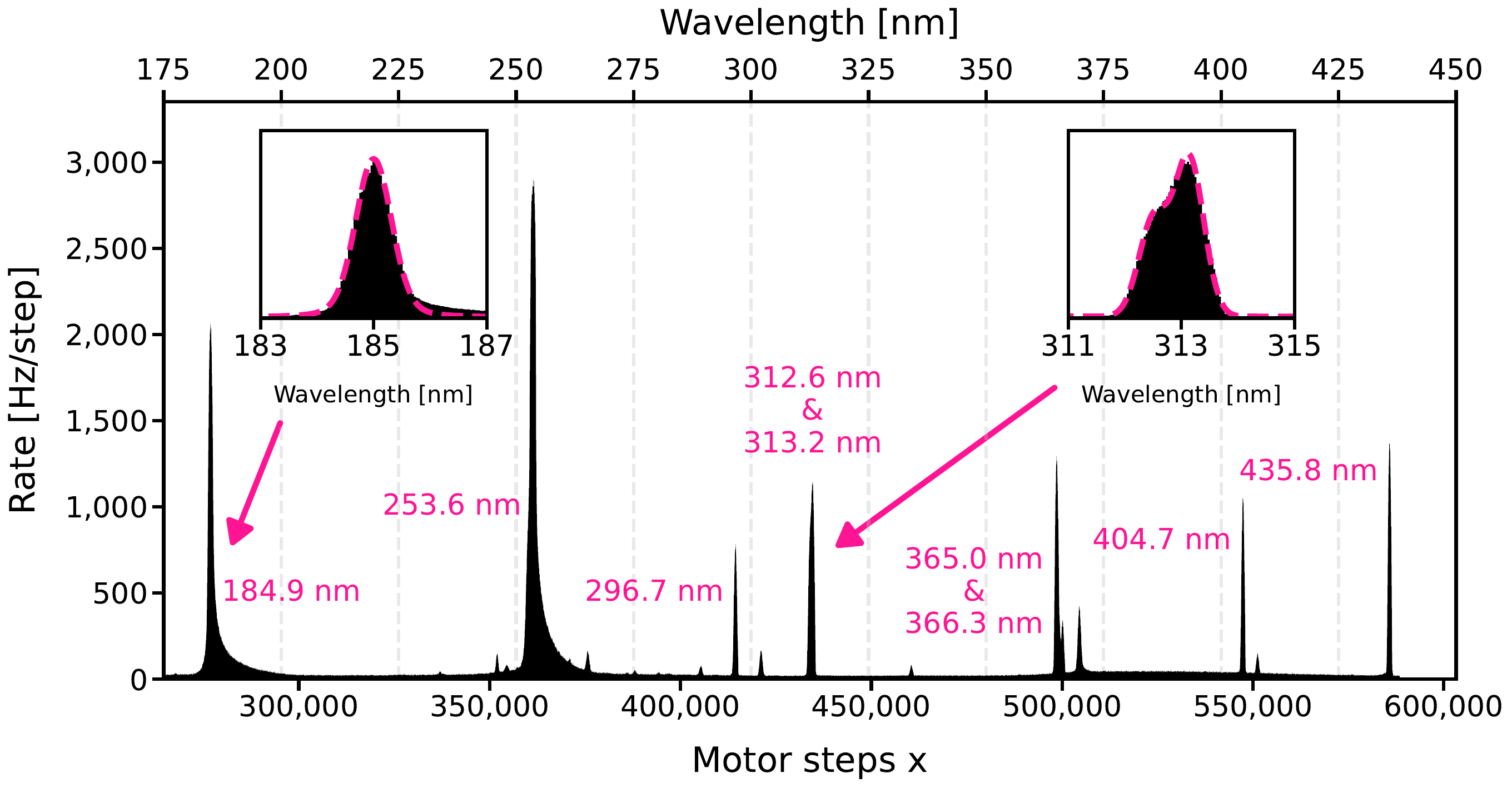}
  \caption{Spectrum acquired from the low-pressure Hg discharge lamp, annotated with the wavelengths of known atomic Hg lines. The inset panels display Voigt profile fits to the atomic line at $184.9$\,nm and the doublet of lines at $312.6$ and $313.2$\,nm, after having applied corrections for the DAQ deadtime. The $184.9$\,nm line serves as a calibration standard within the VUV regime of xenon luminescence, while the doublet illustrates our ability to nearly resolve a spectral feature on the subnanometre scale.}
  \label{fig:HgSpec}
\end{figure}

A Hg lamp mentioned above was used to calibrate the wavelength scale of the monochromator. The atomic emission lines visible in Fig.~\ref{fig:HgSpec} were corrected for the DAQ deadtime then fitted with Voigt profile distributions. Long-wavelength tails from coma aberration are visible in the UV region of the spectrum, and are not adequately modelled by the Voigt profiles.
%The spectrum shown in Fig.~\ref{fig:HgSpec} was corrected for the DAQ deadtime, and the atomic emission lines were fit to Voigt profile distributions.
%For the shortest-wavelength peak ($185$\,nm), attaching an exponential tail to the fit function provides an estimate of the systematic uncertainties, \textcolor{red}{$\bf{X}$/$\bf{X}$}\,nm for the mean/FWHM.
%after having applied DAQ deadtime corrections calculated from a separate dataset measuring the minimum time separation between $^{228}$Th triggers.

The fit of the $184.9$\,nm line was used to calculate FWHM resolutions of $1.0$ and $0.7$\,nm for slit widths of $0.25$ and $0.18$\,mm. The narrower slit provides sufficient resolution to distinguish the $312.6$ and $313.2$\,nm doublet of atomic lines, expanded in Fig.~\ref{fig:HgSpec}. This demonstrates that we can measure the spectral shape of xenon excimer emission with nanometre resolution. Attaching an exponential tail to the $184.9$\,nm fit function results in no significant change to the peak wavelength and FWHM, indicating the systematic uncertainty associated with coma aberration is negligible.
%%%%%%%%%%
%Attaching an exponential tail to the $184.9$\,nm fit function provides an estimate of the systematic uncertainty associated with coma aberration, \textcolor{red}{$\bf{X}$\,($\bf{X}$})\,nm for the mean\,(FWHM).
%%%%%%%%%%

\begin{figure}[ht]
  \centering
  \includegraphics[width=0.65\textwidth]{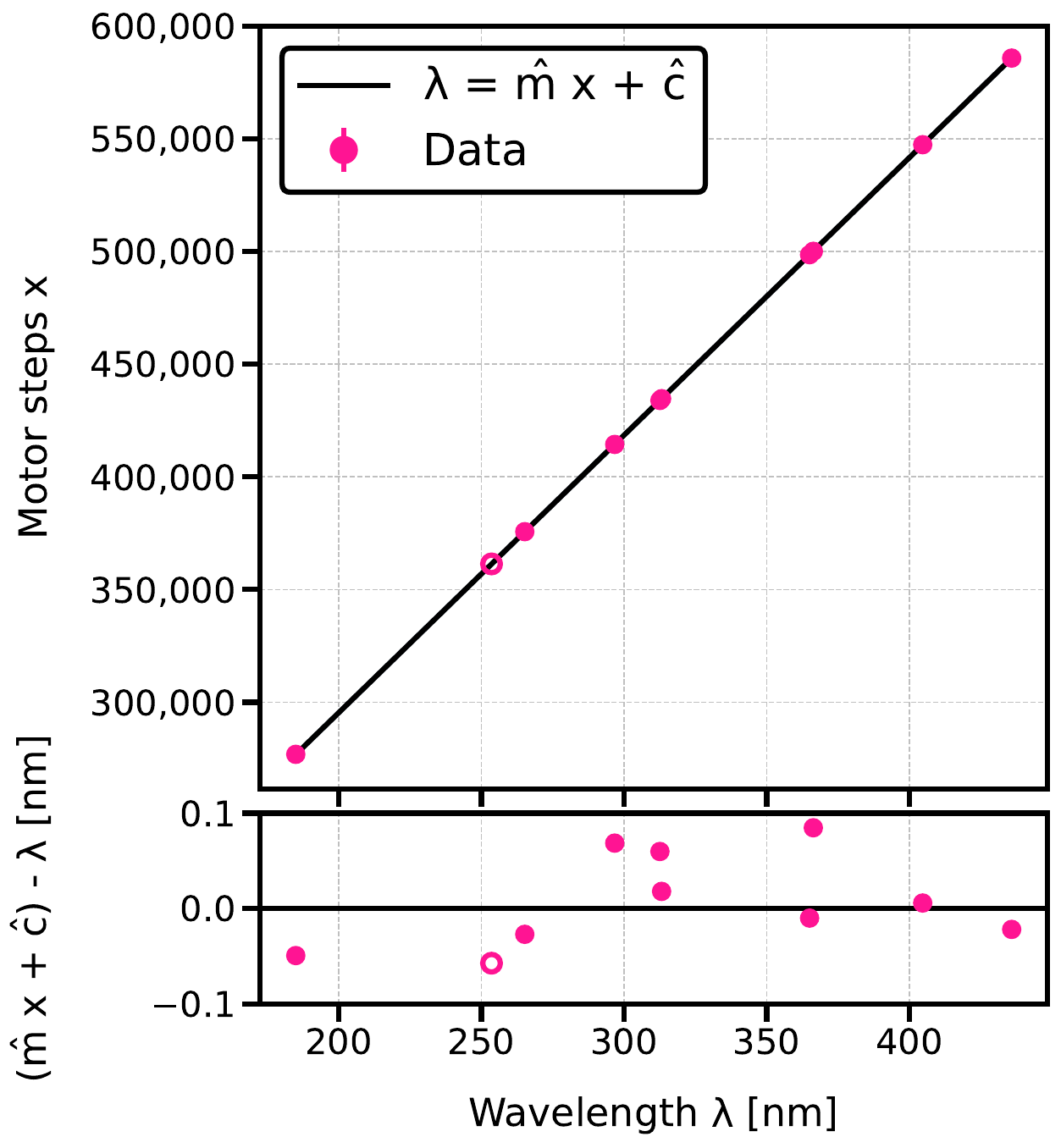}
  \caption{Wavelength calibration of the motor step position conducted using the low-pressure Hg discharge lamp. The $253.6$\,nm line is indicated by an open circle, as it was excluded from the calibration due to the excessive deadtime corrections associated with its high intensity. All datapoints deviate from the linear calibration by $<0.1$\,nm.}
  \label{fig:HgCal}
\end{figure}

A linear fit to the motor position (Voigt profile peak) as a function of mean wavelength of each atomic line is shown in Fig.~\ref{fig:HgCal}. Deviations of the datapoints from the fit are small compared to the resolution of the system, all being $<0.1$\,nm. To ensure stability of this calibration, mercury data were collected repeatedly throughout the month-long period during which xenon-luminescence data were also acquired. Variations in the reconstructed wavelength of the $184.9$\,nm line between calibrations were used to estimate the systematic uncertainty resulting from slight changes in the geometry of the setup, which may be caused by fluctuations in ambient temperature or other factors. The $1\sigma$ uncertainty bounds on the reconstructed wavelength due to these effects are $-0.04$ and $+0.06$\,nm.
%A test was carried out to quantify how a misalignment of the monochromator input from the optical axis of the Hg lamp (and $^{228}$Th source) would affect the calibrations. Hg data were taken with the monochromator moved $\sim\!\pm2$\,mm to the right/left resulting in the recovered mean wavelength of the $185$\,nm line varying by \textcolor{red}{$\bf{X}$/$\bf{X}$}\,nm. These values, together with the absolute deviation of the $185$\,nm line from the fit, are taken as systematic uncertainties on the measurements of xenon luminescence described in the next section.
%by \textcolor{red}{$\bf{X}$/$\bf{X}$}\,nm. 
%The calibration was repeated for $0.25$~mm slits with the resulting slope and intercept deviating by $\bf{<1}$\% from the calibration with the smaller slit.
%There are long-wavelength tails from \textbf{coma aberration} which cannot be modeled by a Voigt profile distribution, thus data from the long-wavelength side of the peak was not included in the fit. The difference between the area of the data and fit is $\bf<5$\%, therefore we use the Voigt profile fits as the apparatus response function that is deconvolved from the xenon spectra in Sec.~\ref{subsec4.4}.
%The FWHM resolution of $\bf{0.67}$~nm for a slit width of $0.18$~mm is calculated from the $185$~nm Voigt profile fit shown in Fig.~\ref{}, and a resolution of $\bf{1.03}$~nm is calculated for $0.25$~mm slits by the same method.

\subsection{Luminescence spectra}\label{subsec3.3}
\subsubsection{Liquid and vapour in equilibrium}\label{subsec3.3.1}
An $\alpha$-decay event from the $2.2$\,bar liquid-vapour dataset is illustrated in Figure~\ref{fig:waveform}. The internal PMT trace contains: a sharp primary scintillation pulse (S1) characterised by the decay times of the singlet and triplet excimer states ($2$--$5$ and $21$--$32$\,ns~\cite{LUXPSD,T1T3A,T1T3B,T1T3C,T1T3D,T1T3E,T1T3F}); a $\sim\!1\,\mu$s delay in which the accompanying ionisation electrons drift through the few millimetres liquid; and an approximately rectangular electroluminescence pulse (S2) with a width corresponding to the time it takes the electrons to travel from the liquid surface to the anode. By contrast, the spectroscopy PMT is poorly coupled to the xenon luminescence and thus it only ever sees single photons. These could be time-coincident with either the S1 or the S2; the black trace in Fig.~\ref{fig:waveform} is a single-photon pulse drawn from the S2 population.

We used $\alpha$ decays for our investigation as they produce large primary scintillation signals (several-hundred-thousand photons) together with many ionisation electrons that generate very large electroluminescence signals (few-million photons). The $^{228}$Th chain also contains $\beta$ decays, but these produce much smaller S1s decreasing the likelihood of detecting a coincident photon with the spectroscopy PMT by an order of magnitude. Additionally, background events with extended tracks of pulses from muons passing through the entire height of the detector are present. To minimize these backgrounds and select pure populations of $\alpha$ decays, we require that each waveform contain a single S1–S2 pair, with both pulses exhibiting appropriate area and width, and separated by the expected electron drift time.
%We used $\alpha$ decays for our investigation as they produce large primary scintillation signals (several-hundred-thousand photons) together with many ionisation electrons that give very large electroluminescence signals (few-million photons). The $^{228}$Th chain also contains $\beta$ decays, but these produce much smaller S1s that give a low probability of detecting a coincident photon in the spectroscopy PMT. There are backgrounds which produce pulses that can be simultaneous mixtures of both types of luminescence: tracks from muons passing through the full height of the detector and $\beta$ decays that are accompanied by emission of a $\gamma$ ray that interacts at another location in the xenon. We reject backgrounds and select pure populations of $\alpha$ decays by: requiring every waveform contain one S1 separated by the drift time from one S2 and using a `k-means clustering'~\cite{kmeans} to define a cut around these populations in (S1,S2)-space. In addition, pulse size and width cuts remove contamination from baseline noise.

\begin{figure}[ht]
  \centering
  \includegraphics[width=\textwidth]{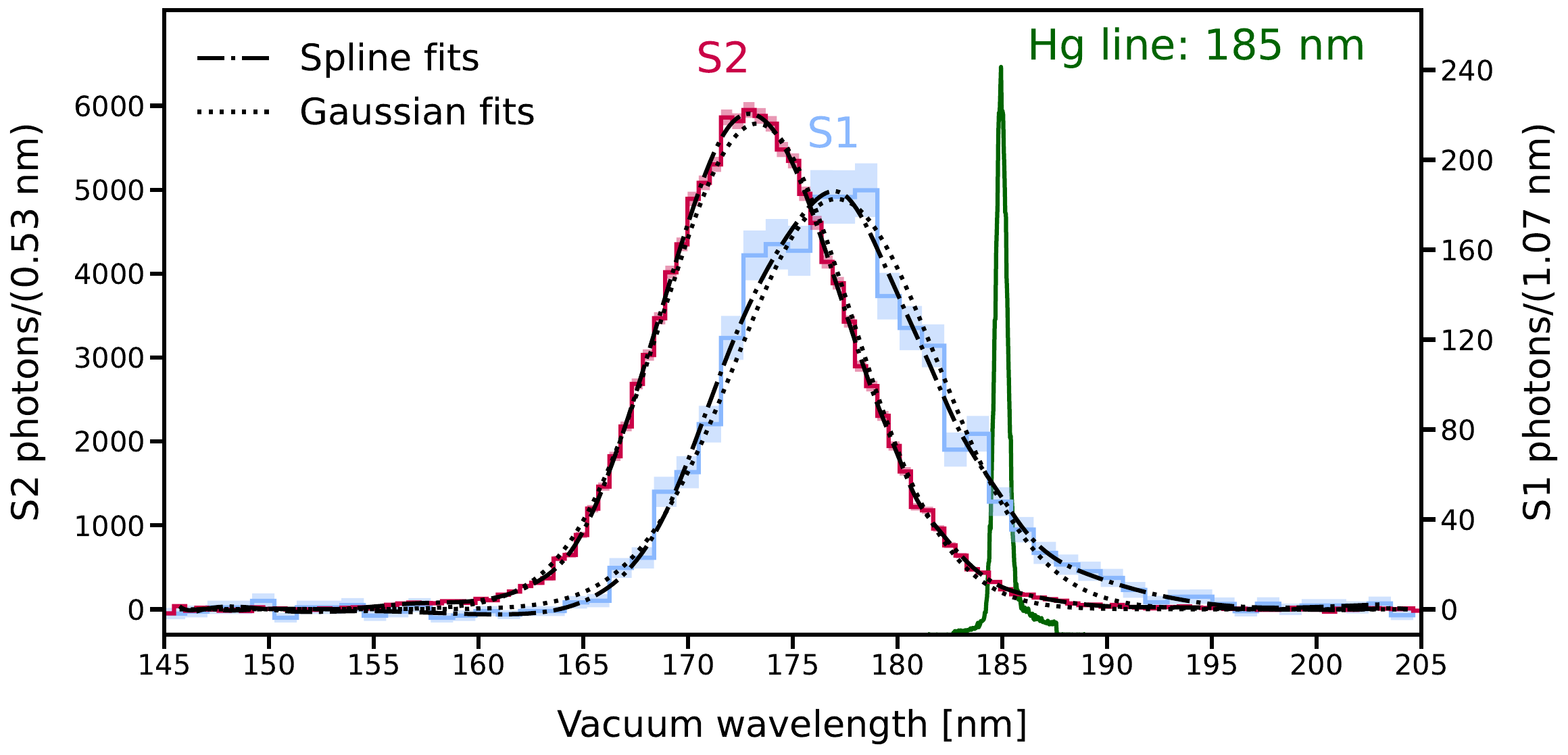}
  \caption{S1 xenon luminescence spectrum for cryogenic liquid and S2 xenon luminescence spectrum for vapour in equilibrium. The data collected at pressures of $1.3$, $1.7$, and $2.2$\,bar are consistent and have therefore been combined in the plot. Each spectrum is parametrised with a Gaussian and a spline. The Hg-lamp $184.9$\,nm calibration line is included on the same plot for comparison. The peak wavelength of the S2 spectrum is shifted toward shorter wavelengths by $\sim\!4$\,nm relative to that of the S1.}
  \label{fig:XeLum}
\end{figure}

The xenon luminescence spectra for liquid and vapour in equilibrium are presented Fig.~\ref{fig:XeLum}. Two spectra are shown, corresponding to photons detected in coincidence with either the S1 or S2 signal. These spectra combine data collected at all three vapour pressures and include the efficiency corrections described in Fig. \ref{fig:efficiency}. Each spectrum exhibits a constant background rate from thermionic dark counts in the spectroscopy PMT occurring within the S1 or S2 time window. Notably, the S1 spectrum is shifted toward longer wavelengths by several nanometres relative to the S2 spectrum, a shift that is comparable to the difference observed between liquid and gaseous phases near the critical point reported in Ref.~\cite{Wahl}.
%Figure~\ref{fig:XeLum} presents the xenon excimer luminescence spectra for liquid and vapour in equilibrium, combining the data at all three vapour pressures and applying the efficiency corrections from Fig.~\ref{fig:efficiency}. 
%%%
%Figure~\ref{fig:XeLum} shows the xenon excimer luminescence spectra for liquid and vapour in equilibrium at $1.7$\,bar, applying the efficiency corrections from Fig.~\ref{fig:efficiency}. All events are required to have a single spectroscopy-PMT photon in coincidence with either the S1 or S2 signal, as illustrated in Fig.~\ref{fig:waveform}. The constant background rate arises from the accidental coincidence of a thermionic dark count in the spectroscopy PMT with the time window of the S1 or S2. The S1 spectrum is shifted toward longer wavelengths by several nanometers relative to the S2 spectrum, a shift comparable to that observed between liquid and gas phases near the critical point, reported in Ref.~\cite{Wahl}.

\begin{figure}[ht]
  \centering
  \includegraphics[width=\textwidth]{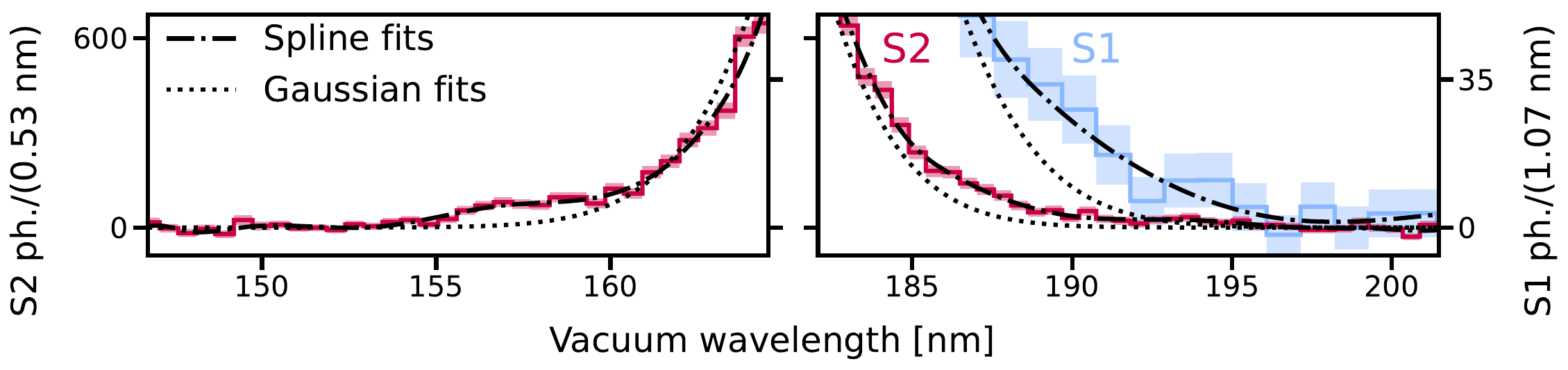}
  \caption{A magnified view of the tails of the S1 and S2 spectra, which are better represented by a spline than a Gaussian. The short-wavelength S1 tail has been omitted from the top-left plot for clarity but can be seen in Fig.~\ref{fig:XeLum}.}
  \label{fig:tails}
\end{figure}

\begin{figure}[ht]
  \centering
  \includegraphics[width=\textwidth]{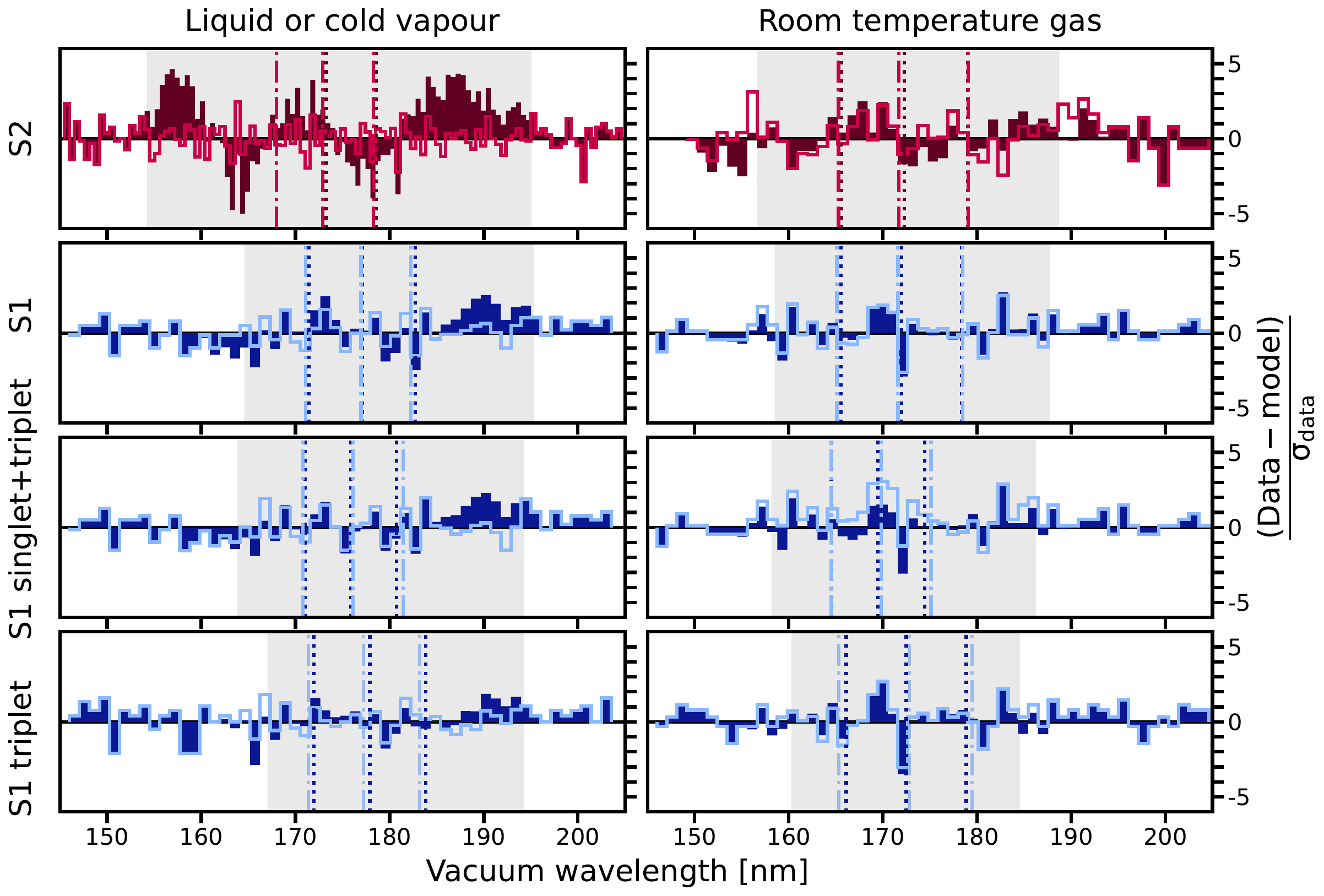}
  \caption{Residuals on Gaussian (solid) and spline (line) reparametrisations of the S1 and S2 spectra. For the liquid-vapour measurements, the residuals of the spline are smaller and more evenly distributed above and below zero across the whole wavelength range, particularly in the tails of the distributions. For the room-temperature gas measurements, both parametrisations accurately model the tails of the distributions. The Gaussian and spline functions for triplet-enhanced S1 data are used to model the triplet component of the combined singlet$+$triplet fit, as described in Sec.~\ref{sec4}. The light grey boxes represent the region in which the fits are greater than one standard deviation above the flat background rate. The peak wavelengths and FWHMs are represented by the dotted (Gaussian) and dot-dashed (spline) vertical lines. In the ``S1 singlet$+$triplet'' plots, these lines correspond to the parameters of the singlet models, which are shifted toward shorter wavelengths compared to those of the triplet models.}
  \label{fig:residuals}
\end{figure}

The spectra are parametrised with two different models: a Gaussian function with a least squares fit to the data, and a spline with the number and position of knots determined by a smoothing condition applied to the reduced chi-squared statistic (implemented in \textit{SciPy's} UnivariateSpline \cite{SciPy}). The spline method was employed to achieve a more accurate representation of the distribution tails, as shown in Fig.~\ref{fig:tails}. Significant improvements are observed on both sides of the S2 spectrum, while the S1 spectrum exhibits improvement primarily on the long-wavelength side. The residuals shown in Fig.~\ref{fig:residuals} demonstrate that the Gaussian fits underestimate both the S1 and S2 tails by several standard deviations --- although they provide a more practical parametrisation which is probably sufficient for most uses.
%The residuals, displayed in Fig.~\ref{fig:residuals}, indicate the Gaussian functions systematically underestimate the S2 tail size by several standard deviations at all pressures.

% \begin{figure}[ht]
%   \centering
%   \includegraphics[trim=0 58 0 0,width=\textwidth]{1p3bar_0p18mm_residual.pdf}
%   \includegraphics[trim=0 58 0 0,width=\textwidth]{1p7bar_0p18mm_residual.pdf}
%   \includegraphics[trim=0 58 0 0,width=\textwidth]{2p2bar_0p25mm_residual.pdf}
%   \includegraphics[width=\textwidth]{GXe_2p2bar_0p18mm_residual.pdf}
%   \caption{Residuals on the Gaussian (solid) and spline (line) paramaterisations of the S1 and S2 spectra. The peak wavelengths and FWHMs are represented by the dotted (Gaussian) and dot-dashed (spline) vertical lines, which agree to within a few-tenths of a nanometer. For the S2 liquid vapour-measurements, the residuals of the spline are smaller and more evenly distributed above and below zero across the whole wavelength range, particularly in the tails of the distributions. There is no evidence of tails on the S1 spectra and the S2 spectrum in room-temperature gas. The light grey boxes represent the region in which the fits are greater than one standard deviation of the flat background rate.}
%   \label{fig:residuals}
% \end{figure}

The S2 short-wavelength tail extends well beyond the main body of the distribution, creating a subtle bump from $154$--$160$\,nm, a region typically associated with first continuum emission. Although this feature constitutes a small fraction of the total spectrum [$(0.6\pm0.1)$\%], it exceeds the background rate of thermionic dark counts by a $2$--$5\sigma$ margin, and its relative magnitude remains consistent across all three vapour pressures. We investigated whether the bump could be an artifact of the monochromator by searching for a comparable feature on the short-wavelength side of the $184.9$\,nm calibration line; however, no evidence of such a feature was observed.

\begin{figure}[ht]
  \centering
  \includegraphics[width=0.8\textwidth]{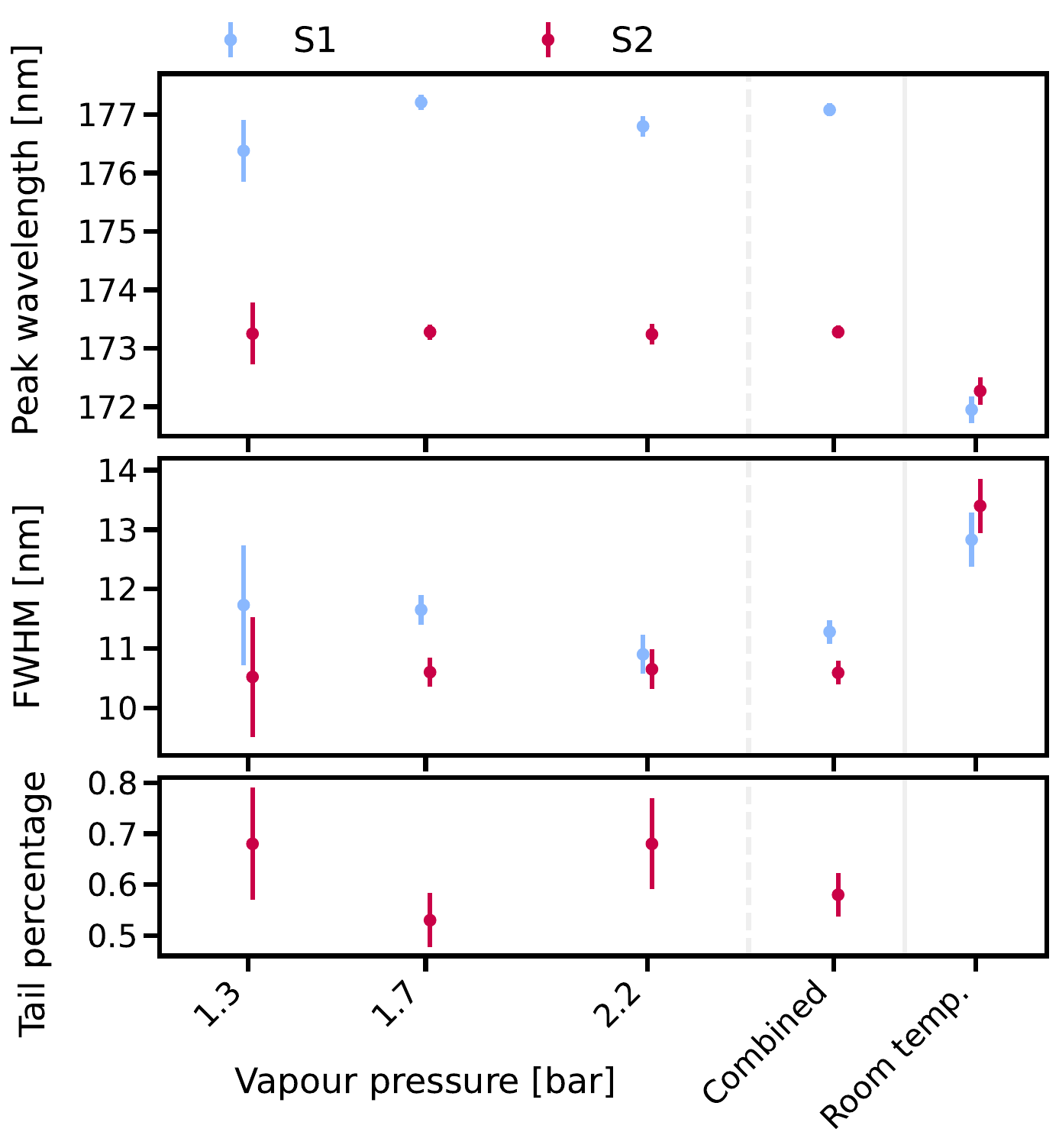}
  \caption{Comparison of the peak wavelength and FWHM measurements extracted from Gaussian fits to the S1 and S2 spectra for liquid and vapour in equilibrium at pressures of $1.3$, $1.7$, and $2.2$\,bar, as well as the combined data. The bottom subplot contains the size of the S2 short-wavelength tail ($154$--$160$\,nm) as a percentage of the total spectrum area. There is good agreement in peak wavelength, FWHM, and tail size across all three vapour pressures. Parameters for the $2.2$\,bar room-temperature gas datasets are provided for comparison: the peak wavelengths are shorter and the FWHMs wider than those in liquid and cold vapour.}
  \label{fig:pressure}
\end{figure}

We searched for potential changes in the S1 and S2 spectra with temperature and pressure by performing Gaussian fits to each dataset separately. Figure~\ref{fig:pressure} demonstrates good agreement in the peak wavelengths and FWHMs for all datasets with liquid and vapour in equilibrium. Furthermore, Kolmogorov-Smirnov (KS) tests indicate the datasets are consistent with the same underlying probability distribution: the S1 and S2 p-values are $0.08$/$0.11$/$0.15$ and $0.45$/$0.18$/$0.06$ for comparisons of the $2.2$--$1.7$/$2.2$--$1.3$/$1.7$--$1.3$\,bar datasets, respectively. Considering the consistency of the peak wavelengths, FWHMs, S2 short-wavelength tails, and KS-test results, we conclude that there are no statistically significant changes in the spectra over the temperature-pressure range studied with liquid and vapour in equilibrium in our experiment. The combined spectrum provides the most precise estimate of the underlying distribution; therefore, the parametrisations of this spectrum are provided for use in optical models of xenon detectors, as discussed in Sec.~\ref{sec4}.

Potential sources of systematic uncertainty arise from our estimations of the MgF$_2$-viewport transmittance and the spectroscopy PMT quantum efficiency (QE). To establish a conservative upper bound on the uncertainty arising from possible discrepancies between the actual transmittance/QE and our estimates, we assume variability of these parameters within the range described in Section~\ref{subsec2.1}. Specifically, we apply a positive tilt of $26$\% and a negative tilt of $14$\% to the overall efficiency curve. This results in a $0.07\pm0.02$\,nm increase or decrease of the Gaussian peak wavelength, while no statistically significant changes were observed in the FWHM or short-wavelength tail. For the S1 spectrum, the statistical uncertainties associated with the fit parameters exceed the magnitude of this effect.
%A potential source of systematic uncertainty originates from our estimation of the spectroscopy-PMT quantum efficiency (QE), which is based on data collected from a different PMT of the same model~\cite{PMTs}. Furthermore, we also lack data on the variability of this parameter across similar samples from this specific manufacturer. However, we have quantified the variability of similar bialkali photocathodes from an alternative supplier~\cite{Hamamatsu} as a $\sim\!14$\% tilt of the QE function within the VUV region.
%However, measurements conducted on bialkali photocathodes from an alternative supplier~\cite{Hamamatsu} indicate $\sim\!14$\% variability in the QE slope within the relevant wavelength range. 
%%%
%we assume that the slope of the efficiency curve could vary by as much as \textcolor{red}{$\bf{\pm14}$\%}. 
%%%
%We currently lack comprehensive data on the variability of QE for bialkali photocathodes produced by this specific manufacturer. 
%%%
%Current LXe experiments have collected extensive data on the variability in detection efficiency of xenon luminescence across various PMTs of a different model.
%An important source of systematic uncertainty for these measurements arises from the QE of the spectroscopy PMT, owing to the fact that we have estimated using data collected from a different PMT of the same model~\textcolor{red}{\cite{}}.

Other systematic uncertainties stem from the reflection and refraction of xenon light on surfaces within the chamber. In most cases, these effects are negligible because only a small fraction of the light is reflected or refracted at the wide angles necessary to reach the monochromator slit ($\sim90$~degrees). We conducted tests of our data to verify each effect is minimal or calculated upper limits on their potential impact using refractive indices and extinction coefficients from Refs.~\cite{refractive_index_liq2,Pt_Au_refractive_index}.
%, or because the variation in reflectivity tends to zero at these angles.
%These effects are highly wavelength-dependent at normal incidence, varying by up to a factor of $\sim\!2$ across the relevant wavelength range. However, at wide angles of $89$--$90$~degrees, which are necessary for light to reach the monochromator slit, these variations decrease to the percent level.

In particular, the geometry of the experimental setup allows a fraction of S2 photons to reflect off the liquid surface toward the monochromator slit, and a smaller fraction of S1 photons to be transmitted through the liquid surface in this direction. As a result, $\sim\!3$\% of S2 photons and $\sim\!0.2$\% of S1 photons detected by the spectroscopy PMT will have undergone reflection or refraction at the liquid surface. For S2s, we confirmed that this effect is small by comparing the spectrum of photons coinciding with the first half of S2 pulses --- where the reflections are present --- with that from the second half of pulses in the $1.7$\,bar dataset. The Gaussian peak wavelength increased by $0.15\pm 0.04$\,nm and the FWHM decreased by $0.16\pm 0.07$\,nm between the two halves of the pulses, while the short-wavelength tail remained a consistent size relative to the overall spectrum. For S1s, we note that the liquid level varies by $\sim\!2$\,mm between the measurements at the three different vapour pressures, and no statistically significant differences were observed between the corresponding spectra.

S1 and S2 light can also reflect off the platinum surface of the source in the direction of the monochromator slit. If these reflections were purely specular, then up to $\sim\!50$\% of S1 photons detected by the spectroscopy PMT could have been reflected in this manner. However, the variation in  reflectivity of platinum across the S1 wavelength range is small at wide angles ($\sim\!1$\%) and is unlikely to significantly affect the shape of the S1 spectrum. Although S2 photons cannot undergo specular reflection from the source into the slit, the surface of the source appears glossy rather than mirror-like, suggesting that contributions from diffuse reflections may be present --- especially at VUV wavelengths. We estimated the magnitude of this effect for S1s and S2s by conservatively assuming that all photons reflect at the angle associated with the greatest variation in photon detection efficiency ($\sim\!11$\% and $\sim\!2$\% across the S1 and S2 wavelength ranges, respectively). Adjusting the spectra accordingly results in no statistically significant changes to the peak wavelengths and FWHMs. (The particular source that we used has a thin layer of gold covering the centre of the disc to prevent the loss of radioactive recoil daughter products. We, therefore, repeated these calculations assuming the surface of the source has the optical properties of gold, and saw the same results.)

\subsubsection{Room-temperature gas}\label{subsec3.3.2}
In room-temperature gas, the S1 data were acquired while the electric field was turned off to prevent the simultaneous generation of S2 electroluminescence, which would otherwise obscure the S1 signal. The S2 data were collected with the electric field established, and the initial $150$\,ns of each pulse were excluded to mitigate overlap with the S1 scintillation. Pure populations of $\alpha$ decays were selected through cuts based on pulse area and width.
%In room-temperature gas, pure populations S1 and S2 signals from $\alpha$ decays were selected through cuts based on pulse area and width. The S1 data were acquired while the electric field was turned off to prevent the simultaneous generation of S2 electroluminescence, which would otherwise obscure the S1 signal. The S2 data where collected while the electric field was powered on, and the initial \textcolor{red}{$\bf{\sim\! X}$}\,ns of each pulse were excluded to eliminate any overlap with the S1 scintillation.
%When the source is in contact with xenon gas, the primary scintillation and electroluminescence signals overlap and form one pulse. To separate out the electroluminescence (S2), we removed the leading \textcolor{red}{$\bf{X}$}\,ns of the pulse, where there is a significant amount of primary scintillation. The cut is driven by the time constant of recombination, which is negligible in the condensed phase but longer than the lifetimes of the excimers in the gas phase  (\textcolor{red}{$\bf{\sim\! X}$}\,ns \textcolor{red}{\textbf{\cite{}}}). To isolate the primary scintillation, we powered down the electrodes so there was no longer any electroluminescence. We then built S1 and S2 spectra by requiring every waveform contain exactly one pulse and selecting populations of $\alpha$ events \textcolor{red}{\textbf{using k-means clustering in one dimension}}.

\begin{figure}[ht]
  \centering
  \includegraphics[height=5.72cm]{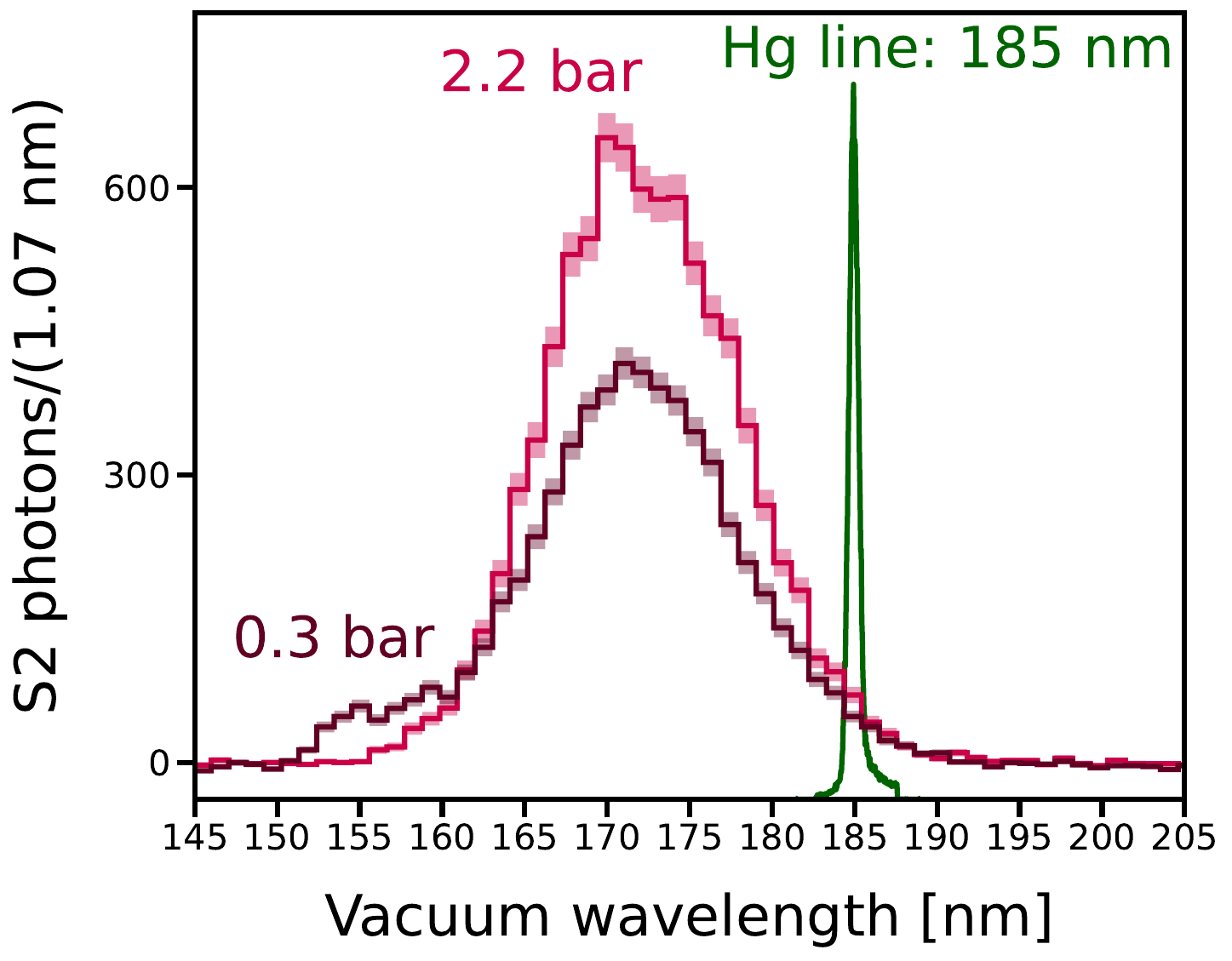}
  \includegraphics[height=5.72cm]{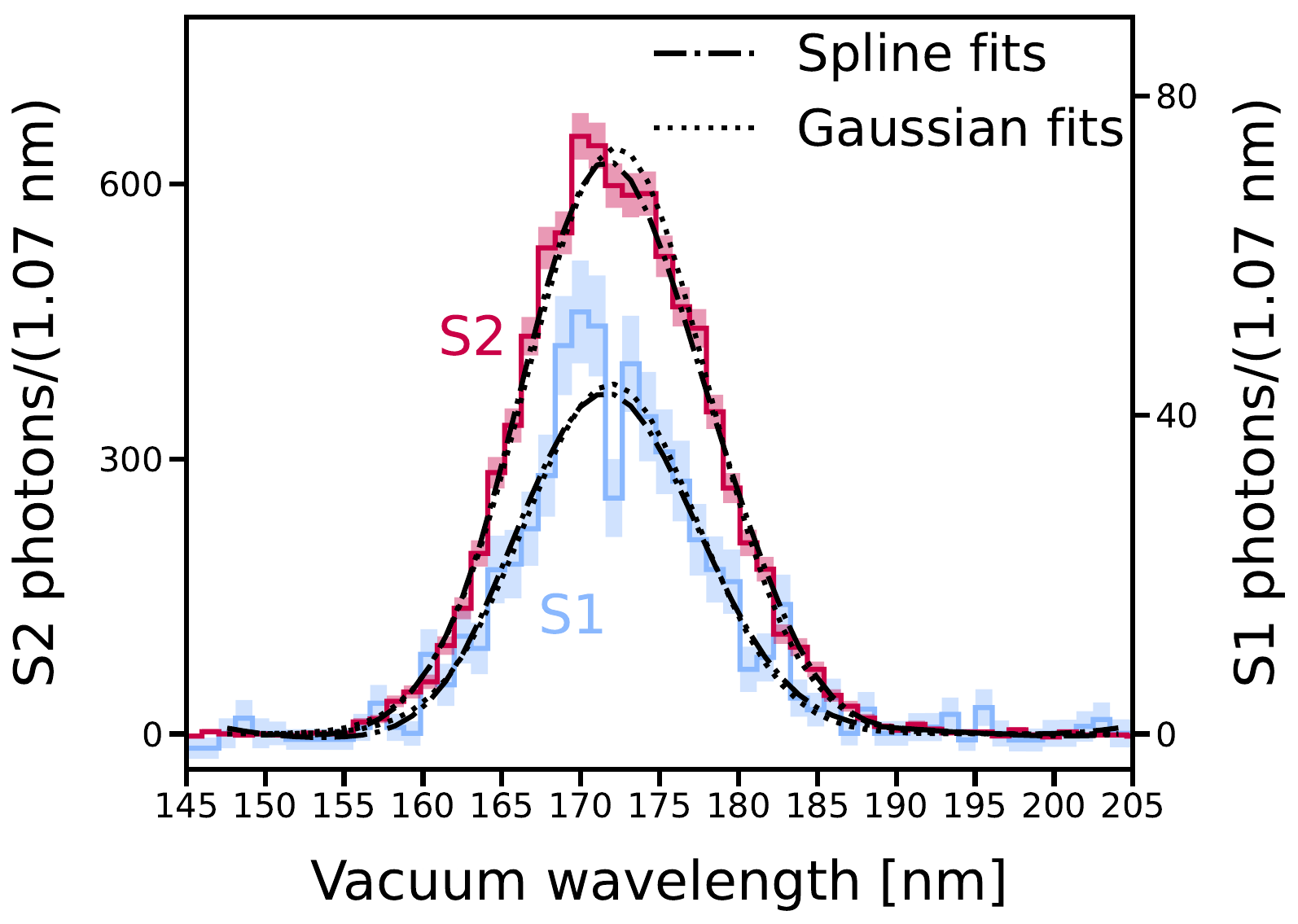}
  \caption{Xenon luminescence spectra taken in room-temperature gas. (Left) A comparison of S2 spectra at $2.2$ and $0.3$\,bar, with the Hg-lamp $184.9$\,nm calibration line overlaid. For both pressures there is a second continuum, while the lower pressure shows an additional first continuum from incomplete collisional de-excitation. (Right) A comparison of S1 and S2 spectra at $2.2$\,bar, with Gaussian and spline parametrisations. The spectra are consistent in their peak wavelength and FWHM, indicating the shift observed between the S1 and S2 spectra in liquid and cold vapour is primarily attributable to the phase of the medium rather than the method of exciting the xenon.}
  \label{fig:XeWarmLum}
\end{figure}

The S2 xenon luminescence spectra taken in $2.2$ and $0.3$\,bar room-temperature gas are presented in Fig.~\ref{fig:XeWarmLum}~(left). The second continuum is evident at both pressures, with a smaller but prominent contribution from the first continuum observed at the lower pressure. This aligns with a previous finding that excimers decay before reaching complete thermalisation via collisions with the surrounding medium at lower pressures~\cite{Marchal}.
%Additionally, in the $0.3$\,bar data, a subtle shoulder is present on the rising edge of the second continuum within the range $162$--$166$\,nm.The shape of the spectrum is consistent from the first to the second half of the dataset, indi

%The size of S2s decreases by 50\% over the length of the $0.3$\,bar dataset, which indicates that impurities are gradually entering the chamber over the timescale of days at this pressure. However, we note that the shape of the $0.3$\,bar spectrum is consistent between the beginning and end of the dataset, suggesting our measurement provides a reasonably accurate representation of the shape of the spectrum. There is not enough data to study the consistency of the subtle bump at 164 nm
%%%
%Additionally, a subtle bump is observed at approximately 164 nm on the rising edge of the $0.3$\,bar second continuum.
%%%
%Figure~\ref{fig:XeWarmLum}~(left) shows the xenon S2 spectra taken in room-temperature gas at $2.2$ and $0.3$\,bar. 
%%%
%It also demonstrates our sensitivity to the spectral region where the S2 short-wavelength tail was observed in cold vapor at higher pressures.
%Although it might be tempting to ascribe this more subtle feature to incomplete thermalisation, we believe this is not the case, as discussed in Sec.~\ref{sec5}.

The $2.2$\,bar S2 spectrum is compared with the S1 spectrum that was recorded at the same temperature and pressure in Figure~\ref{fig:XeWarmLum}~(right). The distributions are aligned at the same peak wavelength, unlike the liquid-vapour measurements, where the S1 spectrum is shifted toward longer wavelengths. This provides additional evidence that the shift is predominantly associated with the phase of the medium, rather than the method of excitation.
%Figure~\ref{fig:XeWarmLum}~(right) compares the $2.2$\,bar S2 spectrum with the S1 spectrum recorded at the same temperature and pressure. 
%%%
%Figure~\ref{fig:XeWarmLum}~(left) shows the xenon S2 spectra taken in room-temperature gas at $2.2$ and $0.3$\,bar. Second continua are visible at both pressures and a smaller contribution from the first continuum is visible at the lower pressure. The presence of the first continuum at the lower pressure is consistent with excimers decaying before they have thermalised with their surroundings. The peak wavelength of the $2.2$\,bar spectrum is $\sim\!1$\,nm shorter than that measured in cold vapour.
%%%
%This indicates that, at the reduced pressure, some excimers decay before they have fully thermalised through collisions with the surrounding medium.
%This may also be an effect related to incomplete thermalisation; however, it lacks the definitive signature of a distinct peak in the region of the first continuum.\textcolor{blue}{\textbf{Additionally, the collisional rate is primarily dependent on the medium's density~\cite{}, which is the same for room-temperature gas at $2.2$\,bar and cold vapor at $1.3$\,bar. Section~\ref{sec5} provides an alternative explanation of how the S2 spectrum may vary due to changes in the Stokes shift with temperature and pressure.}}.

In contrast to the liquid-vapour measurements, both the Gaussian and spline parametrisations of the $2.2$\,bar room-temperature gas spectra accurately represent the distribution tails. Figure~\ref{fig:pressure} compares the Gaussian fit parameters with those from the liquid-vapour measurements. The peak wavelength of the room-temperature gas spectra are $\sim\!1$\,nm shorter than those of the S2 spectra measured in cold vapor, while the FWHM is $\sim\!2$\,nm broader. The reason for these differences between the gas-phase datasets, including the presence of a short-wavelength tail in cold vapour, are discussed in Sec.~\ref{sec4}.
%Figure~\ref{fig:XeWarmLum}~(right) compares the S1 and S2 spectra taken in room-temperature gas at $2.2$\,bar. The S1 spectrum has a second continuum with a peak wavelength that is slightly shorter than that of the S2 spectrum, which is attributed to differences in the ratio of singlet-to-triplet excimers, as described in the next section.
%%%
%The S1 spectrum has a second continuum with a peak wavelength equivalent to that of the S2 spectrum, and no contribution from the first continuum. This consistency indicates the method of exciting the medium --- exciting atoms into the first metastable and resonant states by electron impact versus overexciting atoms through collisions with $\alpha$ particles --- has a subdominant effect on the shape of the spectra. We note that the S2 spectrum is predominantly due to emission from the triplet state, while the S1 spectrum includes a minor contribution from the singlet state ($X$\%) that is expected to shift the spectrum toward shorter wavelengths. However, this effect is not statistically significant in the current data.

A potential source of systematic uncertainty for the $2.2$\,bar room-temperature data is the gradual ingress of contaminants --- observed through the $4$\% reduction in S1 pulse area over the length of the dataset. To evaluate the impact of this systematic effect on the overall shape of the S1 and S2 spectra, we compared the Gaussian peak wavelengths and FWHMs from the first and last halves of the datasets. No statistically significant changes were observed in these parameters. Additionally, KS tests indicate that the data subsets are compatible with the same underlying probability distributions ($p = 0.63$ and $0.99$ for S1s and S2s, respectively).

The presence of a contaminant with a $\sim\!172$\,nm absorption line could explain the low occupancy in the corresponding bin of the S1 histogram. However, the counts in this bin did not decrease significantly over the course of the dataset, indicating that such a contaminate is not likely to have been outgassing over the timescale of days. The most probable contaminants are atmospheric gases (N$_2$, O$_2$, H$_2$O, Ar, and CO$_2$), which do not exhibit isolated absorption lines near $\sim\!172$\,nm~\cite{SpectralAtlas}. While the exact cause of the $\sim\!172$\,nm dip cannot be definitively determined, it is unlikely to represent a genuine feature of the S1 emission spectrum or to be the result of a statistical fluctuation.

We also searched for unanticipated changes in the spectra related to the angle of light scattering on the monochromator grating by comparing the spectrum of photons coinciding with the first or last half of S2s in the $2.2$\,bar room-temperature gas dataset. No statistically significant differences were observed in these spectra.

% For the room-temperature gas data, we examined two additional sources of systematic uncertainty, for which the peak wavelength or FWHM were observed to vary by less than \textcolor{blue}{\textbf{a few tenths of a nanometer}}.
% \begin{itemize}
%   \item The angle of light entering the monochromator was studied by selecting photons that coincide with either the first or second half of S2s in the $2.2$\,bar room-temperature gas dataset. The peak wavelength and FWHM of the Gaussian fits vary by \textcolor{red}{$\bf{0.07\pm 0.12}$}\,nm and \textcolor{red}{$\bf{0.01\pm 0.2}$}\,nm.
%   \item We divided the $2.2$\,bar room-temperature gas dataset into a first and second half to investigate any potential dependence of the spectra on the gradual decrease in purity. The KS test applied to these data subsets indicates they come from the same underlying distribution (\textcolor{red}{$\bf{p = X, X}$} for S1s, S2s), and the peak wavelength and FWHM of the Gaussian fits are consistent with one another.
% \end{itemize}

\subsubsection{Singlet and triplet emission spectra}\label{subsec3.3.3}
Alpha-particle excitation of liquid xenon generates significant quantities of both singlet and triplet excimers, along with ionisation. In the condensed phase, the recombination of ions with electrons produces additional excimers on a timescale that is considerably shorter than the decay times of the singlet and triplet states~\cite{T1T3A}. As a result, the temporal profile of the S1 signal is predominantly determined by the decay characteristics of the excimers. By selecting specific regions within the temporal distribution, it is possible to isolate samples enriched in either singlet or triplet components, thereby enabling a comparison of their emission spectra.
%The excitation of xenon by $\alpha$ decays produces both singlet and triplet excimers in significant proportions~\textcolor{red}{\cite{T1T3A}} which enables the investigation of the spectral differences between these two emission pathways presented in this section.
%%%
%Excitation of xenon by alpha decays results in significant fractions of both singlet and triplet excimers~\textcolor{red}{\cite{LiqAlphaSTFrac1,LiqAlphaSTFrac2}}, which enables an investigation of the spectral differences between these two emission pathways.
%Excitation of xenon by alpha decays results in a significant fraction of singlet excimers~\textcolor{red}{\textbf{\cite{}}}, which is not the case for the $\gamma$ rays and selective excitation methods employed in recent spectroscopic measurements~\textcolor{red}{\textbf{\cite{}}}. This enables an investigate the spectral differences between the singlet and triplet emission pathways.

\begin{figure}[ht]
  \centering
  \includegraphics[width=\textwidth]{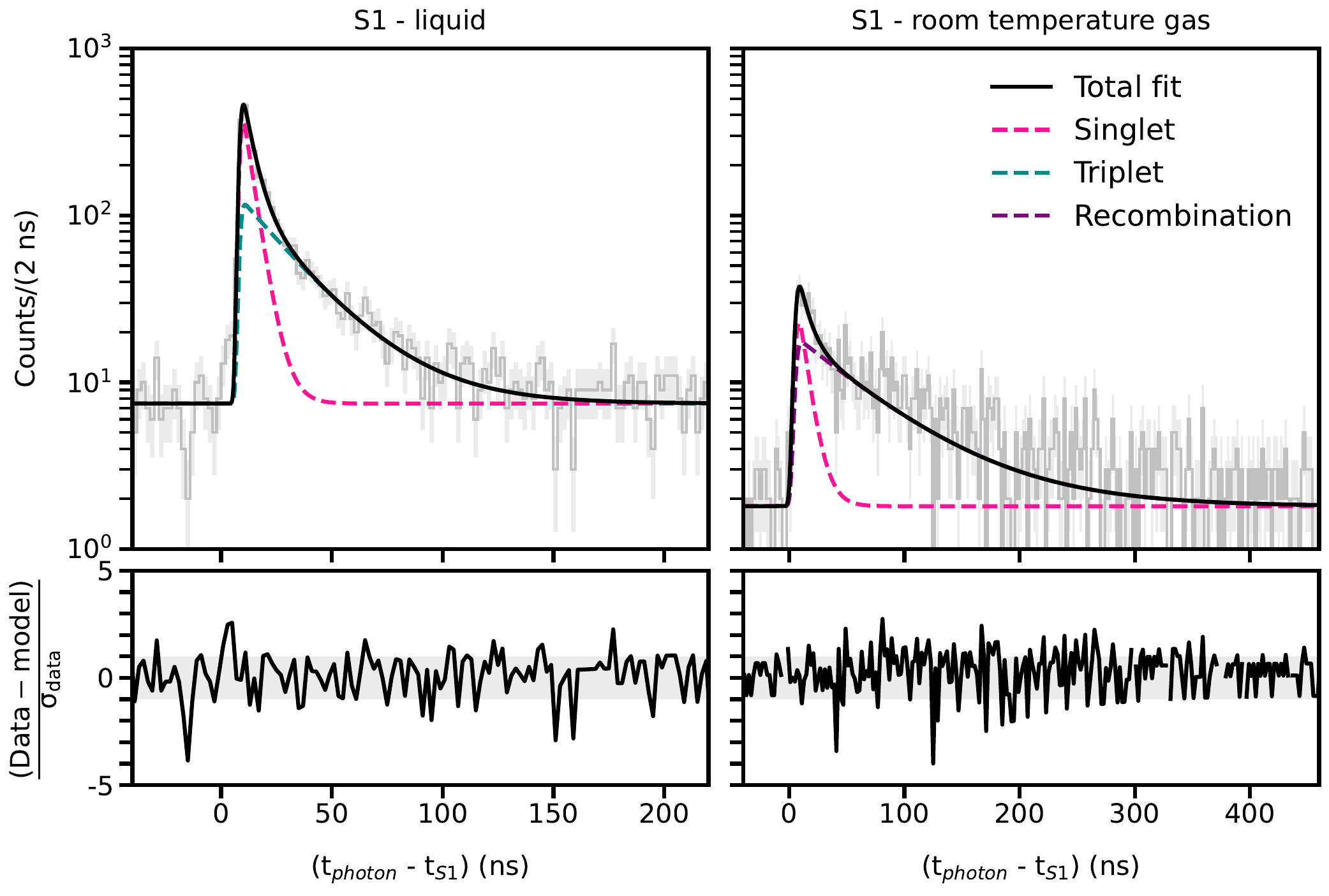}
  \caption{Fit to the distribution of photon detection times recorded by the spectroscopy PMT relative to the S1 pulse start times. The liquid-phase data at all pressures are modelled with a convolution of two exponential decays --- corresponding to the singlet and triplet excimers --- with a Gaussian instrument response function. The room-temperature gas data are fitted by the same functional form, with the singlet decay time constant fixed to the value obtained from the liquid fit; the second exponential component captures the characteristic timescale associated with electron-ion recombination processes.}
  \label{fig:singlet-triplet-fit}
\end{figure}

The photon detection times recorded by the spectroscopy PMT relative to the S1 pulse start times are shown in Fig.~\ref{fig:singlet-triplet-fit}. For decays occurring in the liquid phase, the temporal distribution can be modelled as the convolution of two exponential decays with a Gaussian instrument response function:
$$Y(t)=B+\sum_{i=s,t\,\textrm{or}\,r} C_i \exp\biggl(\frac{\sigma^2}{2\tau_i^2}-\frac{t-t_o}{\tau_i}\biggr)\biggl(1+\erf\biggl(\frac{t-t_o}{\sigma\sqrt{2}}-\frac{\sigma}{\tau_i\sqrt{2}}\biggr)\biggr)\, ,$$
where $\tau_{s,t}$ are the singlet and triplet decay times, $C_{s,t}$ are the corresponding normalisation coefficients, $\sigma$ is the standard deviation of the response function, and $t_o$ is the timing offset between the PMT channels. Fitting the liquid data across all pressures yields decay times of $\tau_{s}=4.9\pm 0.6$\,ns and $\tau_{t}=2
6.8\pm 2.1$\,ns, which are consistent with previously reported values of these parameters, $\tau_{s}\sim2$--$5$\,ns and $\tau_{t}\sim21$--$32$\,ns~\cite{T1T3A,T1T3D,T1T3E,T1T3F,LUXPSD,T1T3C,T1T3B}, respectively.
%The singlet-to-triplet intensity ratio (\textcolor{red}{$\bf{0.82_{-0.11}^{+0.14}}$}) differs from an earlier measurement utilising $\alpha$ decays to excite liquid xenon~\cite{T1T3A}; however, our measurement has low systematic uncertainty in part due to the nearly constant detection efficiency across all wavelengths.

The room-temperature gas measurements were taken at zero electric field where recombination of ions with electrons occurs for as long as several microseconds~\cite{BraisPMTs}, extending the S1 signal well beyond the decay times of the excimers. Without prior knowledge of the ratio of singlet-to-triplet emission resulting from recombination, we cannot precisely attribute different portions of the temporal distribution to these states. However, we can estimate the amount of recombination relative to singlet emission by modelling the data using the same function as for the liquid, with the triplet parameters ($C_t$ and $\tau_t$) replaced by parameters that approximate the recombination process ($C_r$ and $\tau_r$). Fitting the room-temperature gas data to this function yields decay times of $\tau_{s}=8.3\pm 2.4$\,ns and $\tau_{r}=71\pm 7$\,ns.
%and a singlet-to-recombination intensity ratio of \textcolor{red}{$\bf{0.25_{-0.08}^{+0.05}}$}.
%%%
%The room-temperature gas measurements were taken at zero electric field where recombination of ions with electrons occurs over several tens-of-nanoseconds, lengthening the S1 pulse beyond the timescale of the singlet and triplet excimers. These data were modelled using the same function as for the liquid, with the triplet parameters ($C_t$ and $\tau_t$) replaced by parameters characterising the recombination process ($C_r$ and $\tau_r$). 

\begin{figure}[ht]
  \centering
  \includegraphics[width=0.8\textwidth]{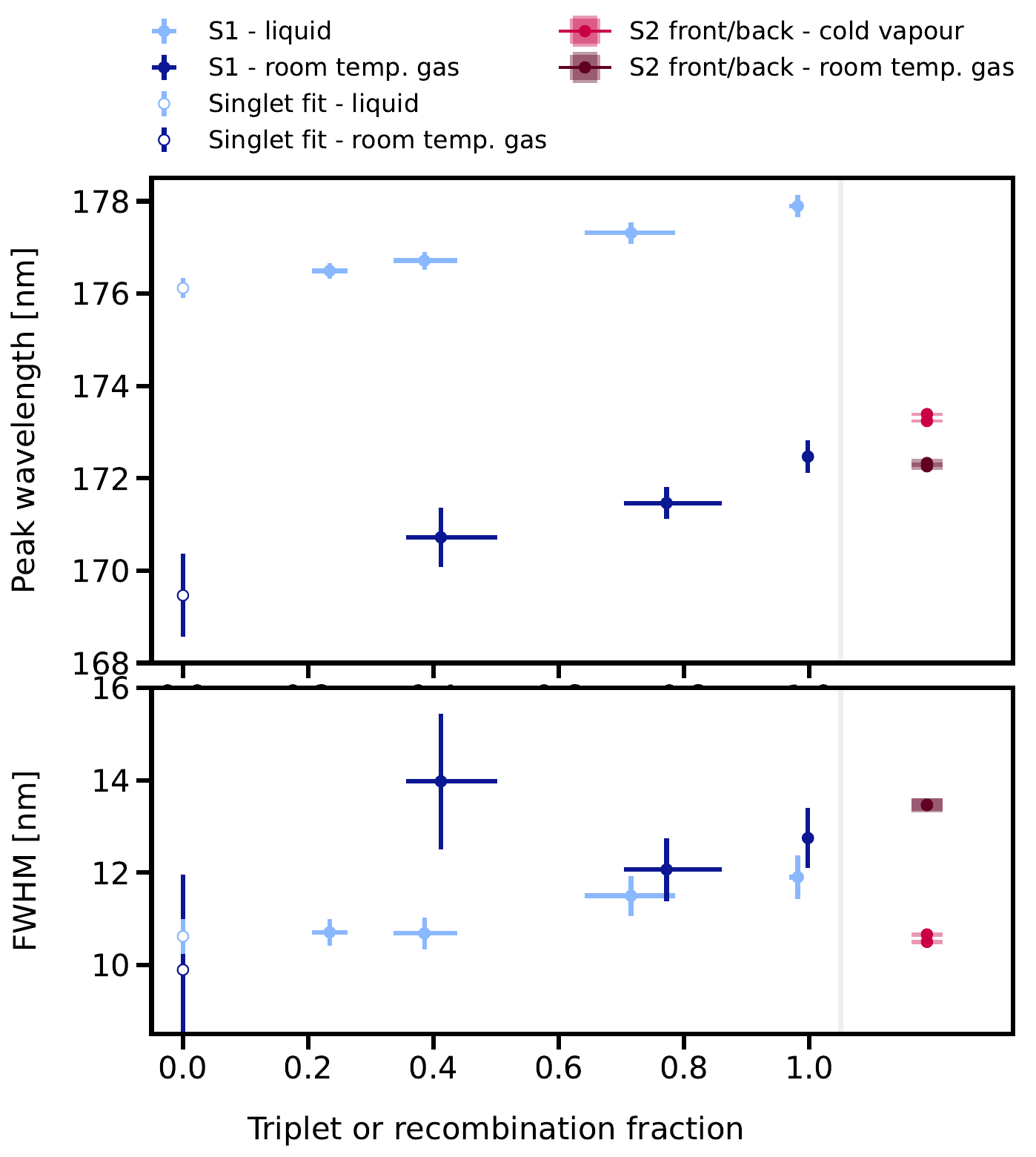}
  \caption{Peak wavelength and FWHM parameters derived from Gaussian fits to S1 spectra with varying singlet-to-triplet (liquid-phase) or singlet-to-recombination (room-temperature gas) ratios. The spectra are generated by applying different selection cuts to the temporal distributions in Fig.~\ref{fig:singlet-triplet-fit}. The singlet datapoints are extracted from the fits that incorporate distinct singlet and triplet components, shown in Fig.~\ref{fig:singlet-calculation}. Plotted for comparison: relatively small changes in peak wavelength and FWHM observed when selecting photons detected in coincidence with the first or second half of S2 pulses.}
  \label{fig:singlet-triplet-trend}
\end{figure}

By selecting different segments of the liquid and room-temperature gas temporal distributions, we generated spectra with distinct triplet or recombination fractions. The peak-wavelength and FWHM values, obtained by fitting Gaussian functions to these subsets of the data, are plotted in Fig.~\ref{fig:singlet-triplet-trend}. The peak wavelength increases by $1.4\pm0.3$\,nm in the liquid and $1.7\pm0.7$\,nm in the room-temperature gas, from the lowest to highest fractions. The FWHM in liquid also exhibits an increasing trend ($1.2\pm0.5$\,nm), but no statistically significant change was observed in room-temperature gas.
%Figure~\ref{fig:singlet-triplet-trend} presents the peak wavelengths and FWHMs obtained by fitting Gaussian functions to subsets of the data with different triplet or recombination fractions, achieved by selecting different segments of the temporal distributions. 
%%%
%In the liquid phase, the peak wavelength~/~FWHM increase by \textcolor{red}{$\bf{1.3\pm0.3}$}\,nm~/~\textcolor{red}{$\bf{1.3\pm0.3}$}\,nm, from the lowest to highest triplet fractions, \textcolor{red}{$\bf{0.23}$--$\bf{0.98}$}. In room-temperature gas, the peak wavelength increases by \textcolor{red}{$\bf{2.0\pm0.8}$}\,nm from a triplet fraction of \textcolor{red}{$\bf{0.38}$--$\bf{1.00}$}, while the FWHM does not show a statistically significant trend.

%The statistical uncertainties of these parameters are represented by the heights of the shaded rectangles.

For comparison with the trends observed in the S1 data, Fig.~\ref{fig:singlet-triplet-trend} illustrates the spectral changes observed when selecting photons that coincide with either the first or second half of S2 pulses. In cold vapour, the S2 fit parameters exhibit variations that are an order of magnitude smaller than those observed for S1 pulses: the peak wavelength and FWHM shift by $0.15\pm 0.04$ and $-0.16\pm 0.07$\,nm, respectively. In room-temperature gas, no statistically significant changes in the S2 fit parameters were observed. Based on this observation, we conclude that selecting different segments of waveforms does not introduce unanticipated systematic effects that bias the S1 trends shown in Fig.~\ref{fig:singlet-triplet-trend}.
%We also note that the S2 parameters for room-temperature gas are consistent with the emission predominantly originating from the triplet state.
%~($\bf{0.07\pm 0.12}$\,nm)
%~($\bf{0.01\pm 0.21}$\,nm)
%For the liquid and vapour data, the variation S2 fit parameters are an order of magnitude less than those observed for S1s: the peak wavelengths increase by \textcolor{red}{$\bf{0.15\pm 0.04}$\,nm} and the FWHMs decrease by \textcolor{red}{$\bf{0.16\pm 0.07}$\,nm} for the cold-vapour~(room-temperature gas) data. 
%The variations in peak wavelength~/~FWHM found by selecting photons coinciding with the first or last half of S2s are plotted on Fig.~\ref{fig:singlet-triplet-trend} for comparison with the singlet-to-triplet trends observed in the S1 data. 
%as indicated by Ref.~\textcolor{red}{\textbf{\cite{}}}, which reported evidence of an efficient transfer towards the triplet state at sufficiently high pressure.
%This conclusion is corroborated by a prior study of xenon excimer formation following selective excitation of the resonant state~\cite{}, which found evidence of an efficient transfer towards the metastable state at sufficiently high pressure.
%by induced symmetry change collisions

\begin{figure}[ht]
  \centering
  \includegraphics[width=\textwidth]{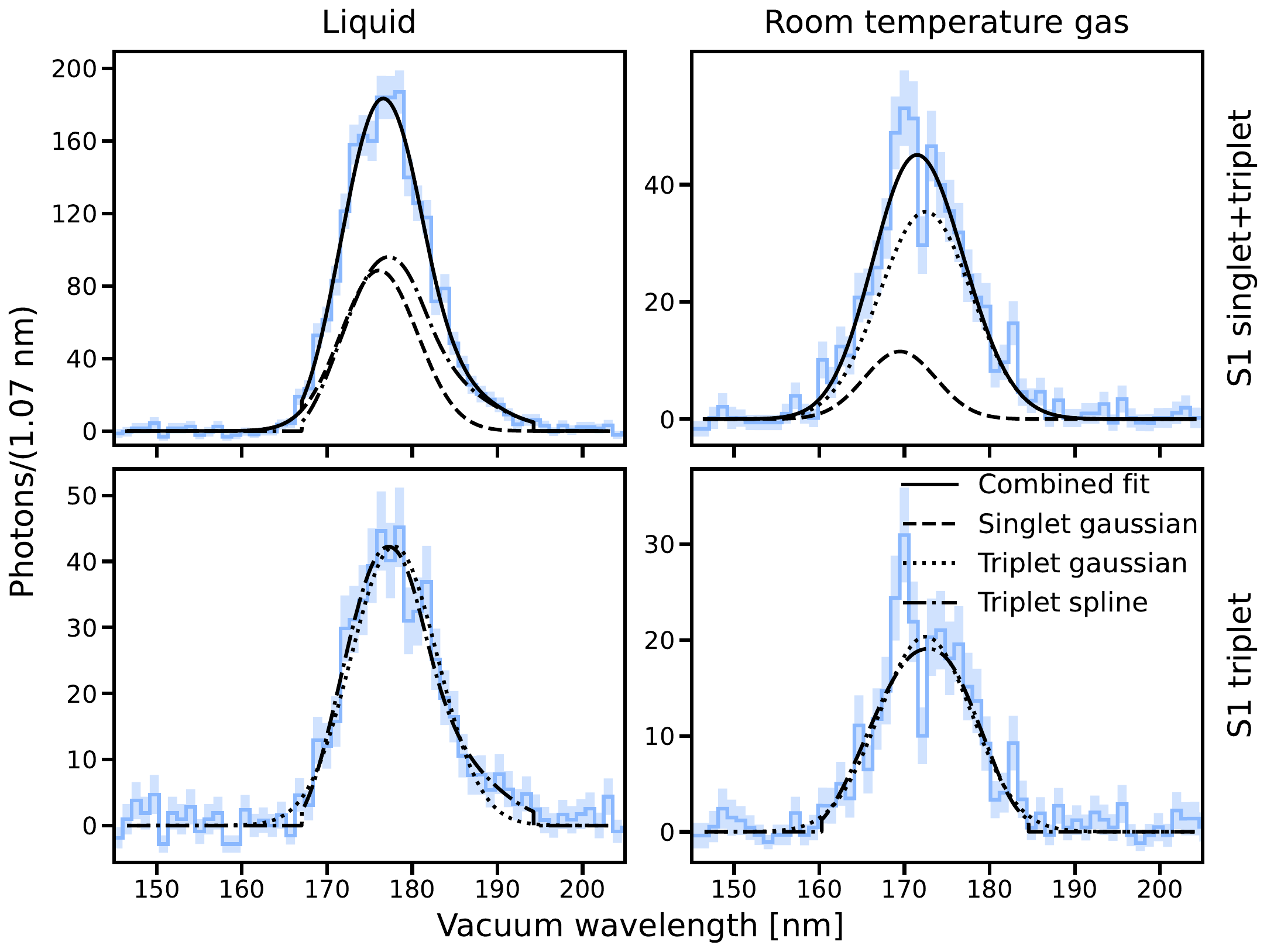}
  \caption{S1 spectra from the liquid and room-temperature gas datasets (top), compared to spectra from a selected subset of the data with an enhanced triplet fraction (bottom). For the room-temperature-gas data we have assumed that recombination results primarily in triplet emission. The triplet-enhanced liquid spectrum exhibits a long-wavelength tail that is more accurately modelled using a spline than a Gaussian. As a result, the corresponding S1 spectrum is fitted using a combination of the triplet spline and a singlet Gaussian. The triplet-enhanced room-temperature gas spectrum is effectively modelled by either a spline or a Gaussian. For simplicity, the corresponding S1 spectrum is fitted using a combination of triplet and singlet Gaussians.}
  \label{fig:singlet-calculation}
\end{figure}

The set of fits used to establish separate singlet and triplet emission models for liquid xenon are presented in Fig.~\ref{fig:singlet-calculation}~(left). The lower left panel shows the spectrum obtained from the subset of data dominated by triplet emission ($98\pm 1$\%). A Gaussian effectively models the central region of the spectrum, while a spline provides a more accurate representation of the long-wavelength tail. The upper left panel shows the spectrum obtained from the entire liquid dataset, which has a triplet fraction of $0.55\pm 0.04$. For these data, a fit function that combines a singlet Gaussian with the previously determined triplet spline effectively models both the centre and tails of the distribution. The singlet peak wavelength and FWHM were varied during the fitting process, while the triplet parameters were fixed at the optimal values obtained in the prior triplet fits. As a result, there is an inherent systematic uncertainty in the singlet parameters arising from the statistical uncertainties associated with the triplet parameters. This uncertainty was evaluated by repeatedly fitting the data with a singlet$+$triplet double Gaussian function, while varying the triplet fraction, peak wavelength, and FWHM based on sampling from the covariance matrices of the triplet fits. The $1\sigma$ uncertainty bounds on the singlet peak wavelength and FWHM due to this effect are $\pm 0.3$\,nm and $\pm 0.5$\,nm, respectively.
%The systematic uncertainty on the singlet parameters, arising from the statistical uncertainty on the triplet parameters, was evaluated by repeatedly fitting the data with singlet and triplet Gaussians. 
%Figure~\ref{fig:singlet-calculation}~(left) presents a set of fits used to establish separate singlet and triplet emission models in liquid xenon.
%A singlet Gaussian model was extracted from these data by performing a combined singlet$+$triplet fit, where the triplet component was modeled with the previously determined spline. 

The process was repeated for the room-temperature gas data in Figure~\ref{fig:singlet-calculation}~(right), assuming recombination results entirely in triplet emission. A Gaussian effectively models both the centre and tails of the triplet-dominated spectrum ($100$\%), eliminating the need for a spline. A singlet Gaussian model is extracted from the entire dataset (triplet~fraction~$=0.80_{-0.04}^{+0.05}$) by performing a combined singlet$+$triplet double Gaussian fit. The $1\sigma$ systematic uncertainty bounds on the singlet peak wavelength are $-1.1$ and $+1.9$\,nm, while the bounds on the FWHM are $-1.7$ and $+4.2$\,nm.

\section{Discussion}\label{sec4}
An objective of this work was to provide primary scintillation and electroluminescence spectra that can serve as inputs to optical models for dual-phase xenon experiments. No significant differences were observed between the spectra from the $1.3$, $1.7$, and $2.2$\,bar datasets; therefore, we recommend using the parametrisations derived from the combined data to model scintillation in liquid and electroluminescence in cold vapour. The Gaussian fits effectively model the central regions of the spectra, while the splines provide a more accurate representation of the tails. For room-temperature gas, the Gaussian fits adequately describe the entire spectral range without the need for a more flexible function. Our results are summarised in Table~\ref{tab:parameterisation}.

In certain cases, it may be advantageous to model scintillation by combining distinct singlet and triplet components. Our S1 spectra accurately represent the emission profiles resulting from $\alpha$-particle excitation of cryogenic liquid and room-temperature gas, which have a significant amount of singlet emission. Other methods of exciting xenon result in different singlet-to-tiplet ratios (e.g. $\gamma$ rays, electrons, xenon nuclear recoils, fission fragment, etc.). For these interactions it may be more appropriate to use the separate singlet and triplet emission models, weighted according to the expected ratio for the specific interaction under investigation. For instance, LXe experiments aimed at detecting dark matter interactions with xenon nuclei must account for background radioactivity that interacts with xenon atomic electrons. The nuclear recoils produce excimers with a singlet-to-triplet ratio of $0.27$~\cite{LUXPSD}, for which we can calculate an average S1 wavelength of $177.8$\,nm using the singlet Gaussian and triplet spline. In contrast, interactions involving electrons generate a significantly lower singlet-to-triplet ratio of $0.042$~\cite{LUXPSD} and a slightly shorter mean wavelength of about $178.2$\,nm.
%The Gaussian fits effectively model the central regions of the singlet and triplet spectra, while the long-wavelength tail of the triplet spectrum in liquid is better captured by the spline. 
%Primary scintillation spectra in both liquid and room-temperature gas should be modeled with the separate singlet and triplet Gaussian components, in the ratio expected for the specific interaction being studied (e.g., $\alpha$ decays. electron recoils, xenon nuclear recoils, etc.). These recommendations are summarised in Table~\ref{tab:parameterisation}.

\begin{table}[h]
\caption{Summary of fit parameters used to model the primary scintillation (S1) and electroluminescence (S2) spectra in the liquid and vapour phases, respectively. Gaussian functions effectively model the centre of all spectra while, in some cases, a spline provides a more accurate representation of the distribution tails. The splines are defined as ${S=\sum_{j=1}^{m+4}\beta_jg_j}$, where $\beta_j$ are coefficients and $g_j$ are the truncated power basis functions for 3rd order splines over the knots $t_j$. These parameters are included in the supplementary materials, along with a corresponding implementation of the function $S$ in Python.}
\label{tab:parameterisation}
\begin{tabular}{@{}lllll@{}}
 & Signal & Function & Peak wavelength (nm) & FWHM (nm) \\
\hline
\hline
 \multirow{7}{*}{\shortstack[l]{Liquid\\ \hspace{0.3cm}\&\\Vapour}} & \multirow{2}{*}{S1} & Gaussian & $177.1\pm0.1_\mathrm{stat} \pm0.1_\mathrm{sys}$ & $11.3\pm0.2_\mathrm{stat} \pm0.0_\mathrm{sys}$ \\
  & & Spline & \hspace{1.2cm}- & \hspace{1.1cm}- \\
  \cmidrule(lr){2-5}
  & S1 singlet & Gaussian & $176.1\pm0.2_\mathrm{stat} \pm0.3_\mathrm{sys}$ & $10.6\pm0.4_\mathrm{stat} {_{-0.5}^{+0.6}}{}_\mathrm{sys}$ \\
  \cmidrule(lr){2-5}
  & \multirow{2}{*}{S1 triplet} & Gaussian & $177.9\pm0.3_\mathrm{stat} \pm0.1_\mathrm{sys}$ & $11.9\pm0.5_\mathrm{stat} \pm0.0_\mathrm{sys}$ \\
  & & Spline & \hspace{1.2cm}- & \hspace{1.1cm}- \\
  \cmidrule(lr){2-5}
  & \multirow{2}{*}{S2} & Gaussian & $173.28\pm0.02_\mathrm{stat} {_{-0.1}^{+0.2}}{}_\mathrm{sys}$ & $10.59\pm0.03_\mathrm{stat} {_{-0.2}^{+0.0}}{}_\mathrm{sys}$ \\
  & & Spline & \hspace{1.2cm}- & \hspace{1.1cm}- \\
\hline
\hline
  \multirow{4}{*}{\shortstack[l]{Room-temp.\\Gas}} & S1 & Gaussian & $172.0\pm0.2_\mathrm{stat}\pm0.1_\mathrm{sys}$ & $12.8\pm0.5_\mathrm{stat} \pm0.0_\mathrm{sys}$ \\
   \cmidrule(lr){2-5}
   & S1 singlet & Gaussian & $169.5\pm0.9_\mathrm{stat} {_{-1.1}^{+1.9}}{}_\mathrm{sys}$ & $9.9\pm2.1_\mathrm{stat} {_{-1.7}^{+4.2}}{}_\mathrm{sys}$ \\
   \cmidrule(lr){2-5}
   & S1 triplet & Gaussian & $172.5\pm0.4_\mathrm{stat}\pm0.1_\mathrm{sys}$ & $12.8\pm0.7_\mathrm{stat} \pm0.0_\mathrm{sys}$ \\
   \cmidrule(lr){2-5}
   & S2 & Gaussian & $172.27\pm0.06_\mathrm{stat} \pm0.1_\mathrm{sys}$ & $13.4\pm0.1_\mathrm{stat} \pm0.0_\mathrm{sys}$ \\
\hline
\end{tabular}
\end{table}

The small but discernible differences observed between the singlet and triplet emission spectra can be understood through the isolated dimer potential energy curves of Fig.~\ref{fig:xelevels}. The singlet and triplet curves are nearly identical in shape and have comparable equilibrium positions; however, the triplet curve is shifted toward lower energies by $\sim\!0.1$\,eV~\cite{Formalism1,Formalism2}. As a result, the two components of the second continuum are expected to be very similar in shape~\cite{FranckCondon}, with the triplet spectrum shifted toward longer wavelengths by several nanometres. The exact shift depends on the specific form of the singlet and triplet potentials, for which there are several different experimentally determined parametrisations for low-pressure gas~\cite{Formalism1,Formalism2,nee2000,Karnbach1995,Xe2StarMorsePotential,XeGroundState}. It also depends on the shape of the ground state potential, which is difficult to predict in the steeply rising region at low internuclear distances~\cite{xiaowei2020}. Our singlet and triplet measurements in room-temperature gas are consistent with those reported in Ref.~\cite{SingletTriplet}. However, we acknowledge that the unexplained feature observed at $172$\,nm may influence our results by a small amount and, therefore, the observed agreement could be somewhat coincidental. Our singlet and triplet measurements conducted for the liquid phase are the first of their kind.
%However, we acknowledge that the unexplained feature observed at $172$\,nm may influence our results, and therefore, the observed agreement could be coincidental. 

Future theoretical work would need to consider the influence of the surrounding medium on the potential energy surfaces to better understand the $\sim\!4$\,nm shift to longer wavelengths of the liquid spectrum compared to that of the vapour. The binary potential energy curves in Fig.~\ref{fig:xelevels} neglect multiparticle interactions, which play a much more significant role in the condensed medium compared to the gas. Additionally, incomplete thermalisation of excimers in the vapour could contribute to the observed shift between phases. This effect would be more significant in our study at pressures of a few bar, where the average interval between atomic collisions is several-hundred-picoseconds, compared the previous study near the critical point~\cite{Wahl}, where the interval would be an order of magnitude shorter. The referenced study observed a similar shift between phases, attributed primarily to the effects of multiparticle interactions.

The triplet emission spectrum in liquid has a peak wavelength $\sim\!3$\,nm longer than the high-resolution measurement of Ref.~\cite{Fujii}, in which $\gamma$-rays were used to excite cryogenic liquid xenon, resulting predominantly in triplet emission. The total S1 spectrum in the liquid is consistent in peak wavelength with the earlier measurement of Ref.~\cite{Jortner}, although our spectrum $\sim\!1$\,nm narrower. This measurement employed an $\alpha$-decay source to excite cryogenic liquid, which would produce a singlet-to-triplet ratio similar to our $^{228}$Th $\alpha$-decay source.
%%%%%%%%%%
%\textcolor{blue}{\textbf{However, that measurement was performed at a temperature ($\sim\!168$\,K) closer to the triple point, and while our results were consistent over the range of temperatures studied here ($170$--$179$\,K), there could be unanticipated variations in the emission spectrum at lower temperatures approaching the solid-vapour equilibrium line. The total S1 spectrum in the liquid is consistent with the earlier measurement of Ref.~\cite{Jortner}, which was performed at a temperature $\sim\!160$\,K and employed an $\alpha$-decay source to excite cryogenic liquid.}}
%%%%%%%%%%
% --- which includes contributions from both singlet and triplet emission in the ratio produced by $\alpha$ decays ---
% Is Co-60 really almost all triplet
% Is the Fuji calibration good enough?
%However, it is important to note that our measurements are the only ones conducted under ultra-high purity conditions, following purification with a zirconium getter. There are several impurities that could absorb on the shorter wavelength side of the emission spectrum () shifting it toward longer wavelengths.
% What are the impurities and does this make sense
% Our warm gas measurements are about the same amount shorter, is the calibration off

Our measurement of the S2 spectrum in cold vapour agrees in peak wavelength with the previous measurement conducted under the same conditions~\cite{Murayama}, although our spectrum is several nanometres narrower. The S2 spectrum in room-temperature gas is consistent with the corresponding measurement in Ref.~\cite{Jortner}, as well as the measurement in gas near the critical point~\cite{Wahl}. The latter was performed at a comparable temperature but a significantly higher pressure/density ($40$\,bar), suggesting that multiparticle interactions do not play a significant role in shaping the gas-phase potential energy curves within the parameter space explored by these studies.

\section{Conclusion}\label{sec5}
In this article, we presented time-resolved photon-counting spectroscopic measurements of VUV luminescence in liquid and gaseous xenon. The main motivation was to accurately characterise the emission spectra of primary scintillation in the liquid phase and secondary scintillation (electroluminescence) in the gas phase in thermal equilibrium, the conditions prevailing in dual-phase xenon detectors used in particle and astroparticle physics experiments.

The primary scintillation spectrum was excited by $\alpha$-particle interactions in the liquid, while the electroluminescence was generated by electron impact in the gas, from ionisation electrons emitted across the liquid surface. We observed a $\sim\!4$\,nm shift in the peak wavelength of the two spectra, with the S1 emission peaking at $177.1\pm0.1_\mathrm{stat} \pm 0.1_\mathrm{sys}$\,nm and the S2 emission peaking at $173.28\pm0.02_\mathrm{stat} {_{-0.1}^{+0.2}}{}_\mathrm{sys}$\,nm.

We have also been able to distinguish the singlet and triplet components of xenon emission for the first time in the liquid phase. We calculated a $\sim\!2$\,nm shift in the peak wavelength between the two spectra, with the singlet emission peaking at $176.1\pm0.2_\mathrm{stat} \pm0.3_\mathrm{sys}$\,nm and the triplet emission peaking at $177.9\pm0.3_\mathrm{stat} \pm0.1_\mathrm{sys}$\,nm. This information is important to understand how the S1 spectrum depends on the excitation mode, as the ratio of singlet-to-triplet excimers is influenced by the excitation process.

The measurements were conducted in pure xenon and the setup was maintained in a stable state over several weeks. The calibration is particularly reliable due to the geometry of the setup, which ensured that light from the calibration source followed the same path as that emitted from the xenon, entering the monochromator at the same angle. The primary source of uncertainty in these measurements stems from reflections on the liquid surface and $^{228}$Th source, which we have confirmed to be a minor effect, as described in Sec.~\ref{subsec3.3.1}. These measurements are the most accurate data available to date to inform optical models for liquid xenon-based particle physics studies.
%We believe that the measurements presented in this paper provide the most accurate characterization of the primary scintillation and electroluminescence spectra relevant to particle physics research utilizing liquid xenon TPCs. These are the only measurements conducted in ultra-high-purity xenon, achieved through purification with a hot zirconium getter and maintained in a stable state over several weeks. The calibration is particularly reliable due to the geometry of the setup, which ensured that light from the calibration source followed the same path as light emitted from the xenon, entering the monochromator at the same angle. The primary source of uncertainty in these measurements stems from reflections on the liquid surface and $^{228}$Th source, which we have demonstrated to be a minor effect. These measurements are the most accurate data available to inform optical models for liquid xenon-based particle physics studies.

\bmhead{Acknowledgments}
This work was funded by the UK’s Science \& Technology Facilities Council (STFC) under the Xenon Futures R\&D Project (ST/T005823/1, ST/V001833/1) and HEP Consolidated Grants (ST/S000739/1, ST/W000636/1). Elisa Jacquet acknowledges funding from the Imperial President’s PhD Scholarship. We thank Eddie Holtom (Rutherford Appleton Laboratory, retired) for significant assistance with the mechanical design work. We are also grateful to the staff of the Physics Instrumentation Workshop and of the Particle Physics Mechanical and Electronics workshops for assistance with design and hardware development. Thanks are due to Alex Lindote and Francisco Neves at LIP-Coimbra, Portugal, for significant assistance with data acquisition and analysis software development. We also extend our gratitude to Andrew Stevens for his valuable contributions to the operation of the liquid xenon system. For the purpose of open access, the authors have applied a Creative Commons Attribution (CC BY) license to any Author Accepted Manuscript version
arising from this submission.

%\section*{Declarations}

%\textcolor{red}{Declarations}

\bibliography{sn-bibliography}

\end{document}